\documentclass[aps,reprint,prb,citeautoscript,superscriptaddress,amsmath,amssymb]{revtex4-2} %reprint,
\usepackage{graphicx}% Include figure files
\usepackage{dcolumn}% Align table columns on decimal point
\usepackage{bm}% bold math
\usepackage{hyperref}% add hypertext capabilities
\usepackage{graphicx}
\usepackage[version=4]{mhchem}
\usepackage[T1]{fontenc}
\usepackage[utf8]{inputenc}
\usepackage{multirow}
 \usepackage{booktabs}
\usepackage{amsmath, amssymb} % para soporte matemático
\usepackage{booktabs,hyperref}

\usepackage{xcolor}

\newcommand{\bluemark}[1] {\color{black}{#1}\color{black}\normalsize}

\usepackage{comment}

\begin{document}

\title{Stacking-Controlled Magnetic Exchange and Magnetoelectric Coupling in Bilayer CrI$_2$}

%\title{Stacking- and strain-dependent magneto-elasticity in 2D CrI${_2}$}

\date{\today} 

\author{B. Valdés-Toro}
\thanks{These authors contributed equally to this work.}
\affiliation{Departamento de Física, Universidad Técnica Federico Santa María, Av. España 1680, Casilla 110V, Valparaíso, Chile}

\author{I. Ferreira-Araya}
\thanks{These authors contributed equally to this work.}
\affiliation{Departamento de Física, Universidad Técnica Federico Santa María, Av. España 1680, Casilla 110V, Valparaíso, Chile}

\author{R. A. Gallardo}
\affiliation{Departamento de Física, Universidad Técnica Federico Santa María, Av. España 1680, Casilla 110V, Valparaíso, Chile}

\author{J. W. Gonz\'alez}
\altaffiliation{Corresponding author:}
\email{jhon.gonzalez@uantof.cl}
\affiliation{Departamento de Física, Universidad de Antofagasta, Av. Angamos 601, Casilla 170, Antofagasta, Chile}

\begin{abstract}
We use a first-principles calculations approach to reveal the electronic and magnetic properties of chromium diiodide (CrI$_2$) bilayers and establish a hierarchy of magnetic interactions across stable registries. The monolayer presents a x-stripe antiferromagnetic ground state, while in bilayers the BA$^\prime$ stacking is the global minimum with antiparallel interlayer magnetic alignment. Bilayer configurations strengthen the exchange in the plane by  6 \% to 10 \%, while the exchange between layers is registry-dependent. The symmetry of each stacking configuration allows for anisotropic interactions. Dzyaloshinskii-Moriya terms appear in structures without inversion symmetry, which in this case also generates in-plane polarizations of up to $\sim$ 10 $\mu$C/cm$^2$, resulting in direct magnetoelectric coupling that is absent in centrosymmetric monolayers. Thus, stacking acts both as a selector of exchange anisotropy and as a driver of magnetoelectricity. Our results show that bilayer CrI$_2$ can be mechanically reconfigured through interlayer sliding, with energy differences between stacking orders (25-50 meV/f.u.) that are compatible with experimental actuation. Tunable magnetism and register-dependent polarization offer promising opportunities for novel spintronic devices, where structural transitions can affect both magnetic states and electric dipoles.
\end{abstract}

\maketitle

\section{Introduction}
Magnetic two-dimensional materials combine reduced dimensionality with tunable symmetry, enabling the control of exchange interactions inaccessible in bulk systems~\cite{parkin2015memory,fert2013skyrmions,chumak2015magnon}. Chromium diiodide (CrI$_2$) shows this potential, exhibiting strong coupling between lattice geometry and magnetism~\cite{yu2025sliding,widyandaru2025tunable}; while bulk CrI$_2$ is ferromagnetic, its monolayer adopts an antiferromagnetic ground state~\cite{zhang2020antiferromagnetic,acharya2025first}. Atomically thin CrI$_2$ layers can be obtained using different experimental techniques, such as mechanical exfoliation, annealing, and molecular beam epitaxy~\cite{li2023two, liu2021mechanical, singh2022van}. In this context, few-layer CrI$_2$ offers a particularly sensitive realization of geometry-driven magnetism in van der Waals crystals.

\bluemark{This sensitivity is characteristic of transition metal-based van der Waals magnets, such as chromium halides (CrI$_3$, CrI$_2$) and tellurides (CrTe$_2$), where subtle variations in anion size, oxidation state, and lattice distortions can qualitatively reshape the magnetic ground state~\cite{huang2017layer,armitage2025electronic,leon2020strain}.
Within this diverse landscape, while CrI$_3$  is established as a prototypical two-dimensional ferromagnetic semiconductor with layer-dependent magnetism~\cite{huang2017layer,lauer2025optimizing}, CrI$_2$  exhibits a fundamentally distinct and richer magnetic phenomenology. 
Such contrast arises from differences in electronic structure and lattice geometry: CrI$_{3}$ hosts Cr$^{3+}$ ($d^{3}$) ions on a honeycomb lattice favoring ferromagnetic superexchange, whereas CrI$_{2}$ contains Cr$^{2+}$ ($d^{4}$) ions on a distorted triangular lattice~\cite{li2023two,schneeloch2024helimagnetism}. 
The resulting orbital degeneracy and competing exchange pathways stabilize an intrinsic antiferromagnetic ground state in monolayer CrI$_{2}$~\cite{zhang2020antiferromagnetic,nunez2025single}, in sharp contrast to the ferromagnetism of CrI$_{3}$. 
Recent experiments have successfully isolated atomically thin CrI$_{2}$ via molecular beam epitaxy and mechanical exfoliation~\cite{li2023two,liu2021mechanical}, revealing noncollinear helimagnetic order and signs of multiferroicity linked to symmetry breaking~\cite{schneeloch2024helimagnetism,acharya2025first,zhang2022structural}. 
Consequently, few-layer CrI$_2$  stands out as an ideal platform to explore the interplay between stacking registry, symmetry, and magnetoelectric coupling in van der Waals heterostructures.}

The central question motivating this work is how stacking geometry governs the relevant magnetic energy scales in CrI$_2$ bilayers. In few-layer systems, rotations and translations between layers set the overall symmetry of the system~\cite{ji2023general}, which in turn dictates the allowed anisotropic interactions. Layer separation and orbital orientation adjust both the magnitude and sign of interlayer exchange through modified electronic hopping pathways~\cite{leon2020strain,li2025mechanical,lauer2025optimizing}. Notably, non-centrosymmetric registries activate Dzyaloshinskii-Moriya interaction (DMI) and symmetric exchange anisotropy, interactions forbidden by symmetry in monolayer CrI$_2$~\cite{acharya2025first,pakdel2025effect}.

\begin{figure*}[!]
\centering
\includegraphics[width=0.95\textwidth]{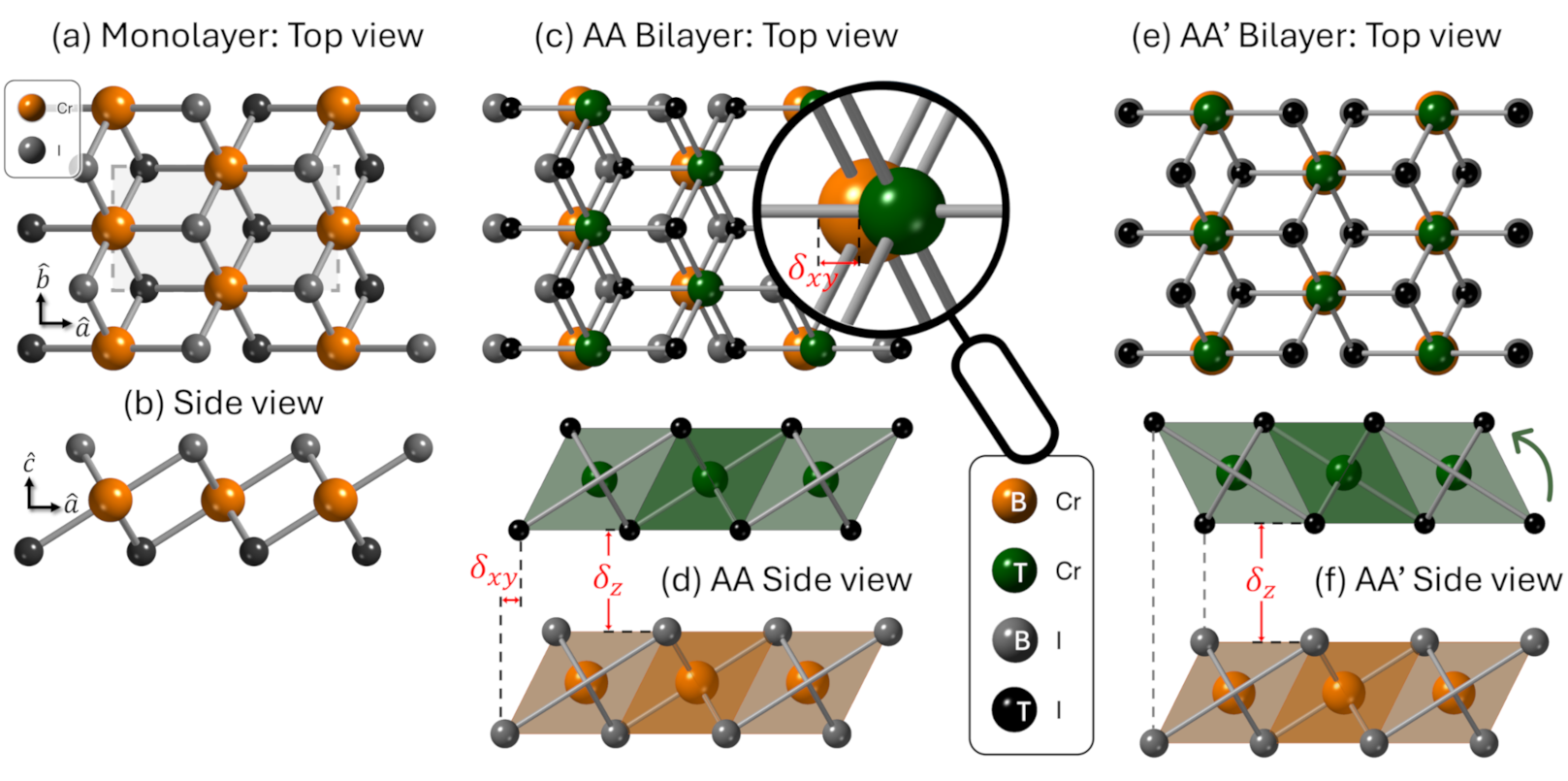}
\caption{
Atomic structures of CrI$_2$ in monolayer and bilayer forms. 
\textbf{(a)} Top view and \textbf{(b)} side view of the relaxed monolayer. 
\textbf{(c,d)} Top and side views of the AA-stacked bilayer. \bluemark{The inset in \textbf{(c)} shows the small in-plane displacement ($\delta_{xy}$) along the $x$ direction that emerges after relaxation, breaking the ideal registry.} 
\textbf{(e,f)} Top and side views of the AA$^\prime$ bilayer, constructed by inverting one layer through the Cr plane followed by a vertical shift. 
In panels \textbf{(c--f)}, different colors distinguish the top (green) and bottom (orange) Cr layers, while I atoms are shown in dark and light gray. 
\bluemark{The side views identify the interlayer distance $\delta_z$, defined between inner iodine planes. }
The Cartesian coordinate system is defined as $\hat{a} \parallel \hat{x}$, $\hat{b} \parallel \hat{y}$, $\hat{c} \parallel \hat{z}$.
}
\label{fig:scheme} 
\end{figure*}

Different registries arise not only by design but also from growth-induced defects. Chemical vapor deposition and subsequent processing often generate folds and controlled buckling that realize twisted or inverted bilayers~\cite{du2021electronically}. Such indirect (mirror-related) registries differ qualitatively from pure translations and have been shown to alter interlayer coupling in other 2D systems~\cite{cortes2018stacking,leon2020strain}. During the growth process, the substrate interactions, thermal gradients, and edge effects can induce stacking faults that propagate across extended domains \cite{akinwande2017review,dai2019strain,ly2017edge,wu2023enhanced}. These defect-mediated geometries open super-superexchange pathways, potentially stabilizing magnetic textures that are inaccessible in ideal crystals. Related principles have enabled stacking-controlled altermagnetism~\cite{gonzalez2025engineering}, stacking ferroelectricity~\cite{yasuda2021stacking}, and valley control~\cite{guo2025valley} in other van der Waals heterostructures.

Recent theoretical studies further demonstrated the crucial role of stacking and symmetry in driving magnetoelectric coupling, supporting the robustness of these mechanisms across different CrI$_2$ phases~\cite{yu2025sliding}. The high sensitivity of the magnetic properties to stacking changes in CrI$_2$ suggests possible applications for reconfigurable spintronic devices. Mechanical folds, slips, or strain actuators can reversibly switch between magnetic states without chemical modification~\cite{dai2019strain,sun2021mechanical}. Based on our findings, we identify a clear hierarchy in the interactions within the system: the intralayer exchange emerges as the primary interaction, significantly influencing the overall magnetic characteristics. The symmetry of the stacking plays a crucial role in dictating the anisotropic terms present in the system, while the interlayer exchange, although weaker, still offers important secondary control over the magnetic properties. This understanding positions CrI$_2$ bilayers as a promising platform for exploring stacking-dependent magnetism and developing mechanically reconfigurable 2D spintronic devices.

\section{Methodology}
We use plane-wave density functional theory as implemented in the \textsc{Quantum ESPRESSO} package (v7.4)\cite{giannozzi2009quantum,giannozzi2020quantum}, employing PBEsol pseudopotentials from PSLibrary\cite{prandini2018precision, dal2014pseudopotentials} with DFT-D3(BJ) dispersion corrections~\cite{grimme2011effect}. 
The plane-wave cutoffs are 60/500~Ry for wavefunctions/charge density. Structural relaxations and self-consistent calculations use $\Gamma$-centered $k$-meshes of $4\times 8\times 1$ and $6\times 13\times 1$, respectively, Marzari-Vanderbilt smearing of 0.01~Ry~\cite{marzari1999thermal}, and a vacuum spacing of 18~\AA{} along $z$.

\begin{figure*}[!]
\centering
\raisebox{1.5cm}{\includegraphics[width=0.3\columnwidth]{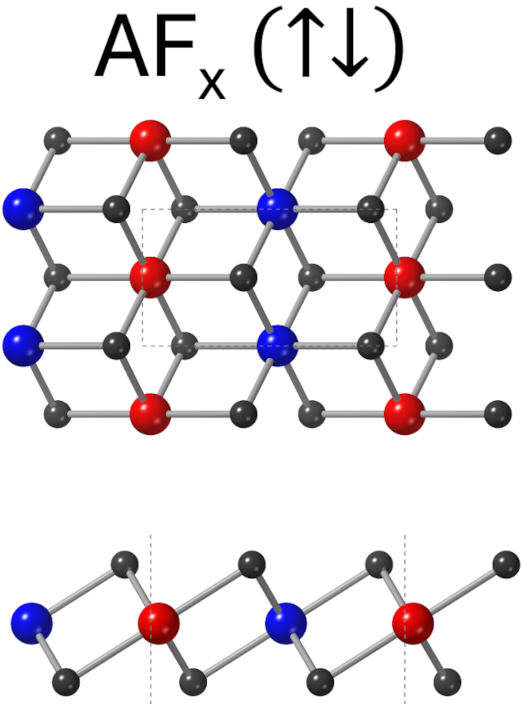}}
\includegraphics[width=0.7\columnwidth]{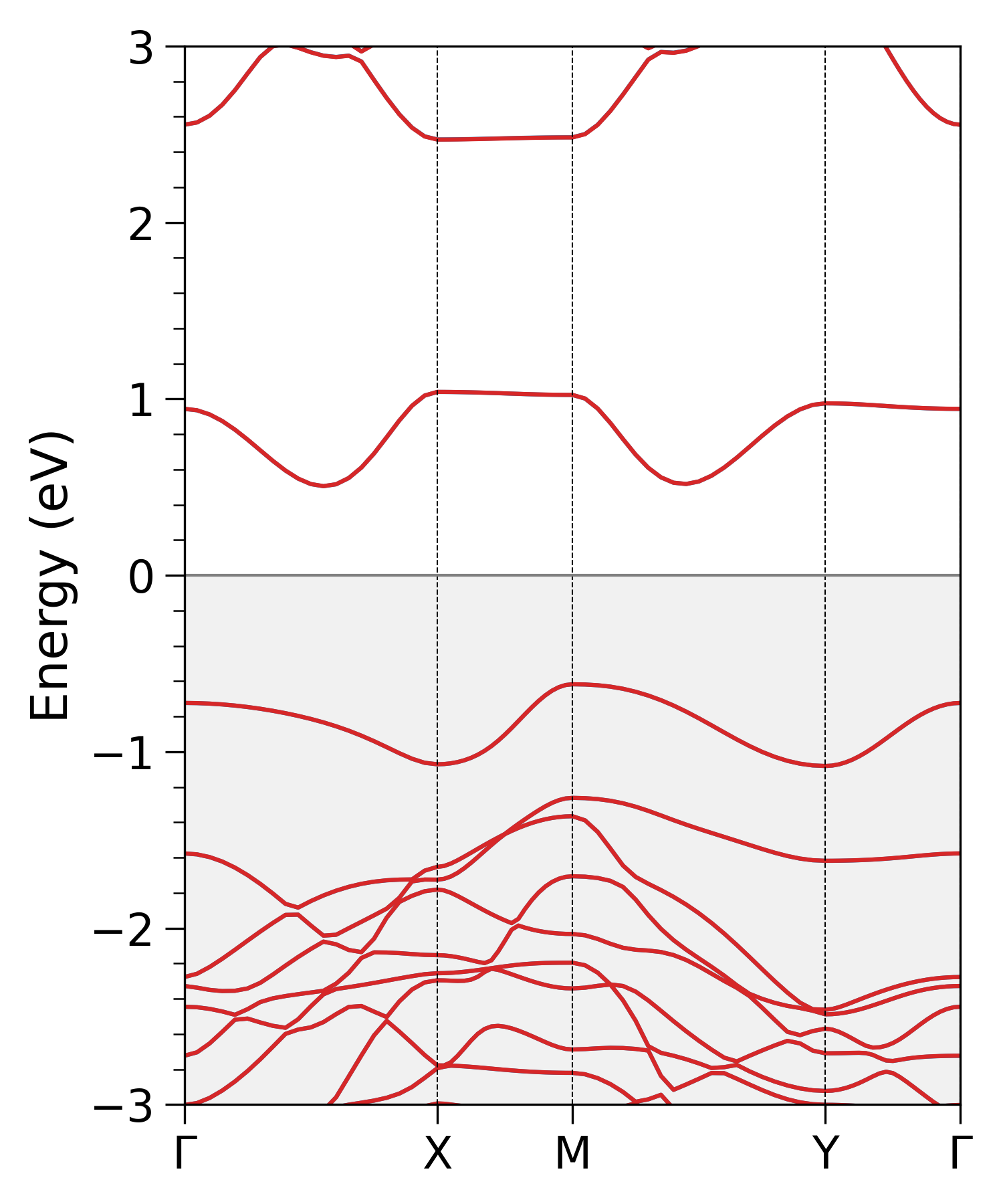}
\raisebox{1.4cm}{\includegraphics[width=0.3\columnwidth]{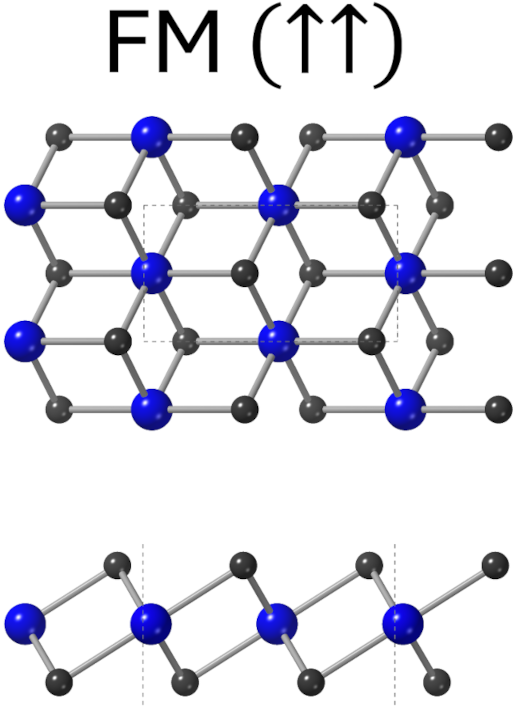}}
\includegraphics[width=0.7\columnwidth]{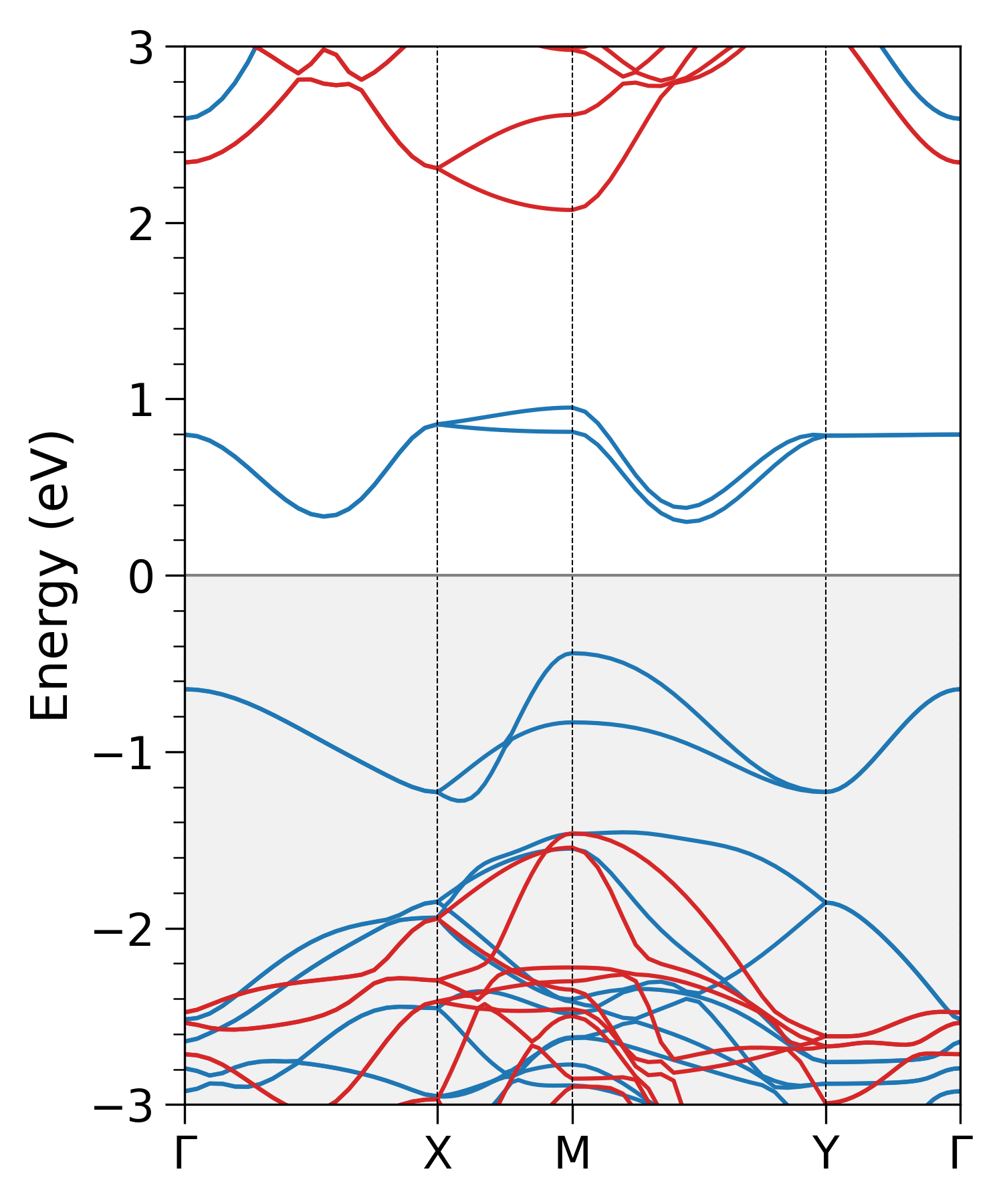}
\caption{Electronic band structures of monolayer CrI$_2$ in the stripe-type antiferromagnetic AF$_x$ \textbf{(left)} and ferromagnetic \textbf{(right)} configurations. Calculations are shown along the high-symmetry path $\Gamma$-X-M-Y-$\Gamma$, where $\Gamma=(0,0)$, X=$(1/2,0)$, M=$(1/2,1/2)$, and Y=$(0,1/2)$ in reciprocal lattice units. The Fermi level is set to zero (represented by a horizontal line).}
\label{fig:ML_bands} 
\end{figure*}

\bluemark{%
In our calculations, we employ the rotationally invariant DFT+$U$ formalism (Liechtenstein formulation), in which the on-site Coulomb repulsion $U$ and Hund’s exchange $J_0$ are treated as independent parameters ($U = 3.1$~eV, $J_0 = 0.1$~eV), yielding an effective interaction $U_{\mathrm{eff}} = U - J_0 = 3.0$~eV. This value lies within the $3.0$--$3.5$~eV range shown to accurately reproduce the electronic structure, magnetic moments, and exchange energetics of chromium halides in benchmark studies against hybrid functionals, at a fraction of the computational cost~\cite{lauer2025optimizing}. Retaining a finite $J_0$, rather than adopting the simplified Dudarev $U_{\mathrm{eff}}$-only limit, provides a more physically complete description of intra-atomic exchange and orbital polarization in the Cr-$3d$ manifold and has been shown to yield a more consistent treatment of spin densities and magnetic energetics in correlated magnetic systems~\cite{ryee2018effect}. This approach is particularly appropriate for Cr$^{2+}$ compounds such as CrI$_2$, where Hund’s coupling can subtly influence exchange pathways and the relative stability of competing magnetic states, and it is fully consistent with recent first-principles studies that successfully captured the helimagnetic ground state of monolayer CrI$_2$ using the same effective interaction scale~\cite{schneeloch2024helimagnetism,li2023two,acharya2025first}.}

We compute magnetic interactions by combining \textsc{OpenMX} (v3.9)~\cite{ozaki2003variationally,ohwaki2014method} and \textsc{TB2J} (v0.9)~\cite{he2021tb2j}. Following a structural relaxation, we perform a self-consistent DFT+$U$ calculation in \textsc{OpenMX} using GGA-PBE, numerical pseudoatomic orbitals, 500 Ry real-space cutoff, and $5\times9\times3$ $k$-mesh with identical $U$ parameters. We perform scalar-relativistic and fully-relativistic calculations including spin-orbit coupling (SOC).
We extract the magnetic exchange tensors $\mathsf{J}_{ij}$ using the magnetic force theorem using the \textsc{TB2J} tool to construct the spin Hamiltonian,
\begin{equation}
H=-\sum_{ij}\mathbf{S}i^{\mathsf{T}}\mathsf{J}{ij}\mathbf{S}_j.
\end{equation}
Without SOC the exchange can be write as, $\mathsf{J}{ij}=J{ij}\mathsf{I}$, and with SOC, $\mathsf{J}{ij}=J^{\mathrm{iso}}{ij}\mathsf{I}+\Gamma_{ij}+\mathsf{A}(\mathbf{D}{ij})$, where $J^{\mathrm{iso}}{ij}$ is isotropic exchange, $\Gamma_{ij}$ is symmetric anisotropy, and $\mathsf{A}(\mathbf{D}{ij})$ encodes Dzyaloshinskii-Moriya interactions. We group crystallographically equivalent bonds into distance shells, reporting shell-averaged couplings: Heisenberg $J{\mathrm{eff}}$ (no SOC) or $J_\alpha=J^{\mathrm{iso}}+\langle\Gamma_{\alpha\alpha}\rangle$, in-plane anisotropy $\Delta J_{xy}=J_x-J_y$, and DMI magnitude $\langle|\mathbf{D}|\rangle$ (with SOC).

\section{Results}
In this section, we present the results of our first-principles calculations. We begin by determining the structural and magnetic ground state of the CrI$_2$ monolayer. Subsequently, we extend our analysis to bilayer systems, investigating the energetic stability of different stacking configurations and their corresponding magnetic and elastic properties.

\subsection{Monolayer: Ground state}
The CrI$_2$ monolayer adopts the typical 1T geometry of transition-metal compounds~\cite{zhang2020antiferromagnetic,gonzalez2024mos2}, with a central Cr plane sandwiched between upper and lower I layers (Fig.~\ref{fig:scheme}), forming edge-sharing CrI$_6$ octahedra. We employ a rectangular $2\times1$ supercell to accommodate competing collinear magnetic orders and evaluate four configurations: ferromagnetic (FM), two stripe antiferromagnetic arrangements (AF$_x$, AF$_y$), and zigzag antiferromagnetic (AF$_z$).

The AF$_x$ stripe order emerges as the ground state, consistent with previous studies~\cite{zhang2020antiferromagnetic,widyandaru2025tunable,acharya2025first}. The magnetic energy hierarchy follows AF$_x$ (ground state) < AF$_z$ (+4.85 meV/Cr) < AF$_y$ (+5.49 meV/Cr) < FM (+7.14 meV/Cr). Structural parameters show minimal magnetic-order dependence: lattice parameters vary by <1\% between AF$_x$ and FM states, while Cr--Cr distances and Cr--I bonds remain nearly identical, reflecting the structural robustness of edge-sharing CrI$_6$ octahedra~\cite{zhang2020antiferromagnetic,acharya2025first}.
We verify the stability of the AF$_x$ ground state across Hubbard $U$ values, finding antiferromagnetic coupling favorable for $U < 6.0$~eV. This sensitivity reflects the superexchange scaling $J \sim t^2/U$ and aligns with established values for chromium halides~\cite{pakdel2025effect,acharya2025first,widyandaru2025tunable,zhang2020antiferromagnetic}. At our chosen $U = 3.1$~eV, the AF$_x$ state lies 7.14~meV/Cr below FM, with local magnetic moments of $\sim 4.2~\mu_B$ per Cr atom.

The AF$_x$ electronic structure reveals a narrow-gap semiconductor with $E_g \sim 1.12$~eV, characterized by anisotropic dispersion and nearly degenerate subbands that reflect the two Cr sublattices in the stripe order. The band edges comprise hybridized Cr-$d$ ($t_{2g}$-like) and I-$p$ states, consistent with previous studies of antiferromagnetic, semiconducting monolayer CrI$_2$~\cite{zhang2020antiferromagnetic,acharya2025first,widyandaru2025tunable}. The proximity of subbands and moderate bandwidth suggests strong strain response, known to tune AFM-FM balance and induce metal-semiconductor transitions in 2D materials~\cite{widyandaru2025tunable,gonzalez2025strain,leon2020strain,gonzalez2021strain}. This electronic foundation establishes the reference for analyzing bilayer stacking effects.

%%%%%

\begin{figure*}[!]
\centering
\raisebox{1.5cm}{\includegraphics[width=0.3\columnwidth]{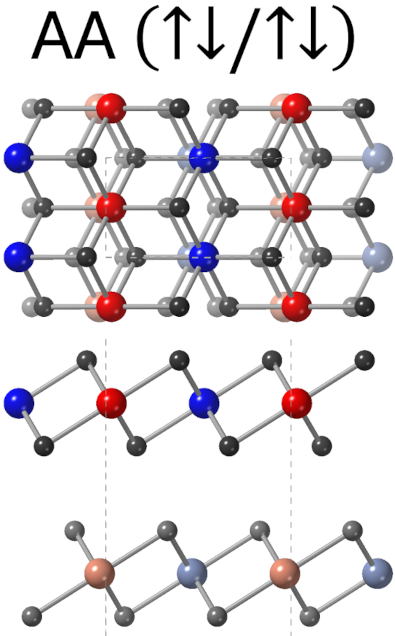}}
\includegraphics[width=0.7\columnwidth]{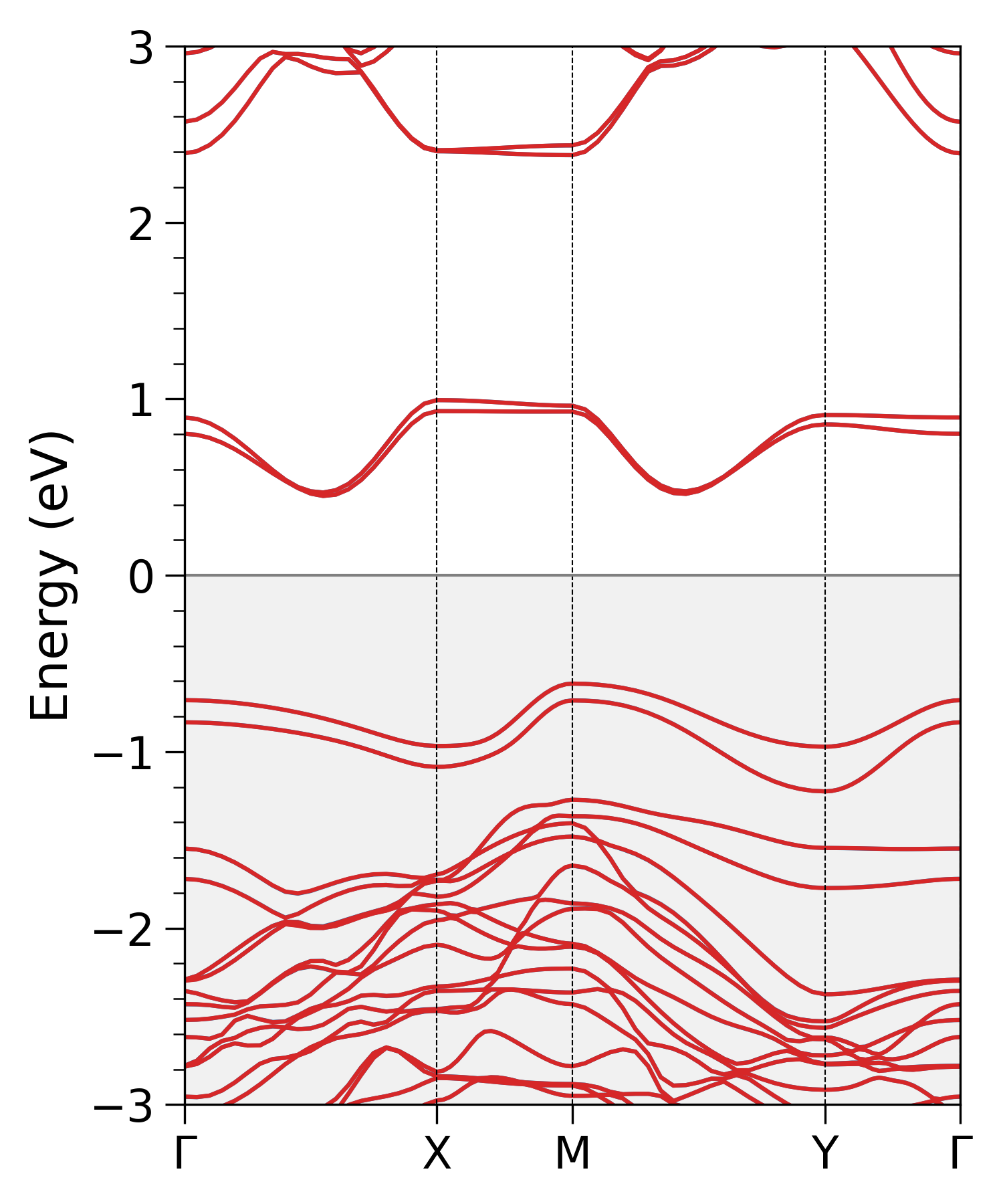}
\raisebox{1.4cm}{\includegraphics[width=0.3\columnwidth]{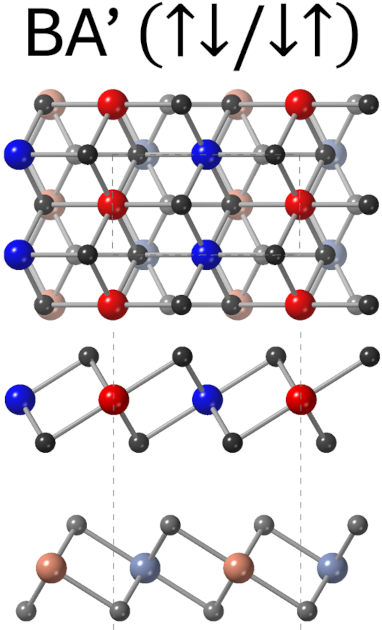}}
\includegraphics[width=0.7\columnwidth]{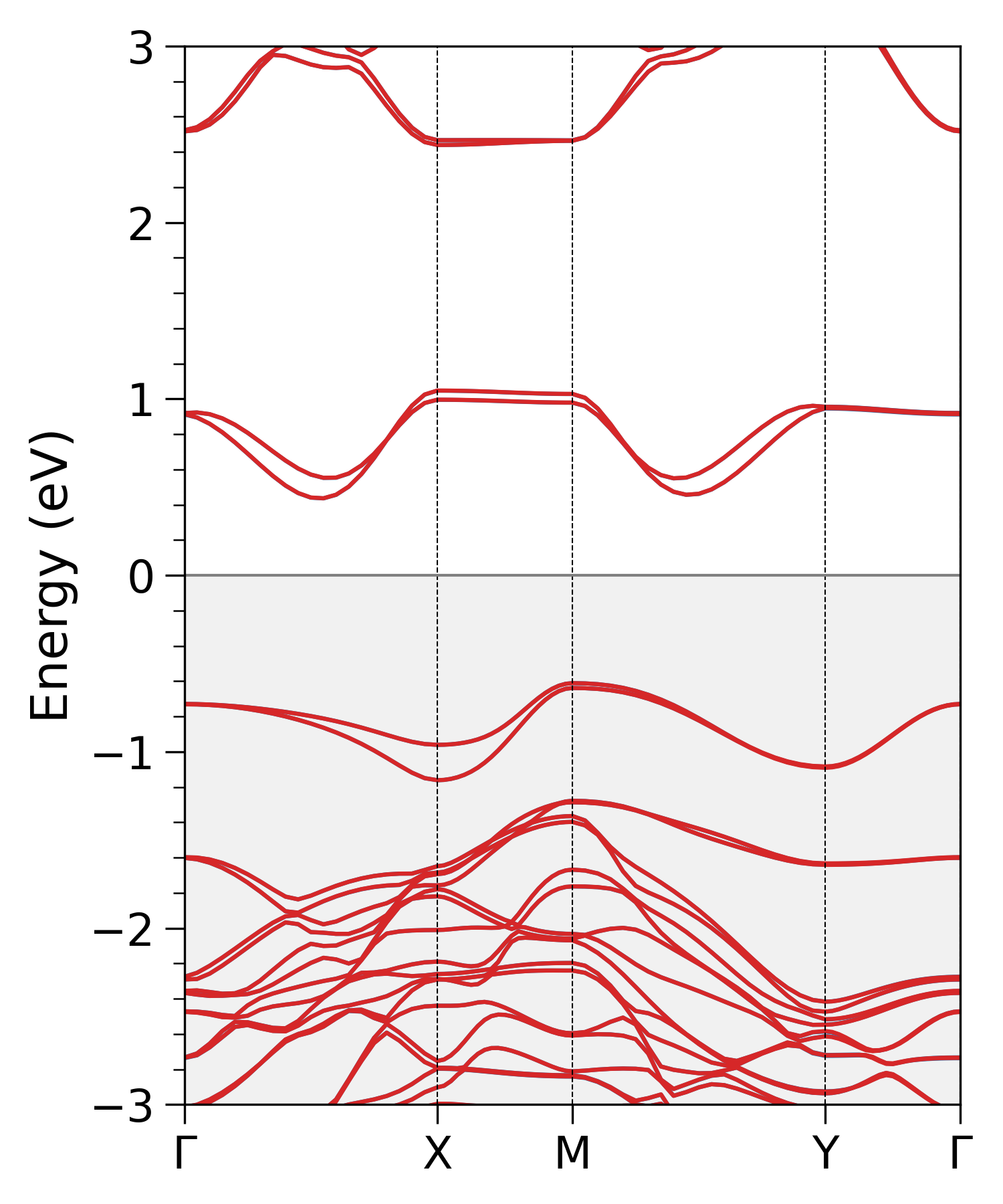}
\caption{Electronic band structures of bilayer CrI$_2$ in the most stable magnetic configurations: direct AA ($\uparrow\downarrow/\uparrow\downarrow$) \textbf{(left)} and BA$^\prime$ ($\uparrow\downarrow/\downarrow\uparrow$) \textbf{(right)}. The calculations are performed along the high-symmetry path $\Gamma$--X--M--Y--$\Gamma$, with $\Gamma=(0,0)$, X=$(1/2,0)$, M=$(1/2,1/2)$, and Y=$(0,1/2)$ in reciprocal lattice units. The Fermi level is set to zero (represented by the horizontal dashed line).}
\label{fig:BL_bands} 
\end{figure*}

\subsection{Bilayer: Ground State}

Motivated by the antiferromagnetic ground state of monolayer CrI$_2$~\cite{zhang2020antiferromagnetic,widyandaru2025tunable} and the strong stacking sensitivity of its properties~\cite{acharya2025first}, we investigate CrI$_2$ bilayer systems constructed from two fundamental stacking families: direct (non-prime) and indirect (prime). The direct stackings (e.g., AA, AB, BA), such as the AA bilayer shown in Fig.~\ref{fig:scheme}(c-d), are generated by creating a translated replica of the first layer. In contrast, the indirect stackings (e.g., AA$^\prime$, AB$^\prime$, BA$^\prime$), illustrated by the AA$^\prime$ bilayer in Fig.~\ref{fig:scheme}(e-f), are constructed through a two-step process: first, the second monolayer is inverted with respect to the out-of-plane axis ($z \rightarrow -z$), an operation equivalent to an in-plane rotation of 180$^\circ$, before being displaced to its final position~\cite{gonzalez2025engineering,cortes2018stacking}. This fundamental difference in symmetry operations is critical, as it dictates which spatial symmetries, such as inversion centers or mirror planes, are preserved or broken in the bilayer system.

Regarding magnetism, we investigate four collinear magnetic configurations for each bilayer stacking by combining intralayer AF$_x$ and FM orders with either parallel or antiparallel interlayer alignments. We denote the spin state of the Cr$_1$-Cr$_2$ pair within each layer using $\uparrow\downarrow$ for the AF$_x$ stripe and $\uparrow\uparrow$ for the FM order.
This approach yields two interlayer-parallel configurations: one where both layers are antiferromagnetic ($\uparrow\downarrow/\uparrow\downarrow$) and another where both are ferromagnetic ($\uparrow\uparrow/\uparrow\uparrow$). It also produces two interlayer-antiparallel configurations, consisting of opposing AFM layers ($\uparrow\downarrow/\downarrow\uparrow$) and opposing FM layers ($\uparrow\uparrow/\downarrow\downarrow$). This set of configurations captures the possible low-energy collinear manifold required to characterize the exchange couplings in CrI$_2$ bilayers~\cite{acharya2025first,widyandaru2025tunable}.

Our analysis of the energies reveals a clear stability ordering, with the BA$^\prime$ family emerging as the global energy minimum. Specifically, the ground state is BA$^\prime$($\uparrow\downarrow/\downarrow\uparrow$), with the companion AFM alignment being nearly degenerate (Table~\ref{tab:D3_per_atom}). The AB$^\prime$ states are the next most stable, lying only a few to tens of meV above the minimum. By contrast, direct (AB/BA) and high-symmetry (AA, AA$^\prime$) stackings are energetically less favorable, ranging from $25-50$ meV/f.u. above the ground state. This energy ordering is consistent with observations in related bilayer magnets such as CrI$_3$, where stacking-dependent energetics also favor specific indirect registries due to optimal interlayer interactions~\cite{leon2020strain,lauer2025optimizing}. Similarly, other 2D magnetic bilayers, such as CrBr$_3$ and CrCl$_3$, exhibit comparable stacking sensitivities, albeit with different ground-state preferences~\cite{li2025mechanical,ghojavand2024strain}.
The interplay between interlayer steric repulsion and structural relaxation drives the energy ordering. For instance, the energetically less favorable AA stacking attempts to alleviate iodine-iodine repulsion by relaxing with a slight lateral shift ($\delta_{xy} \sim 0.43-0.51$~\AA) (Fig.~\ref{fig:scheme}(d)). These structural trends indicate that prime stackings such as BA$^\prime$ and AB$^\prime$, which combine layer inversion with lateral displacement, provide a more effective mechanism to minimize interlayer repulsion and stabilize the bilayer ground state.

\begin{figure*}[!]
\centering
\includegraphics[width=0.8\textwidth]{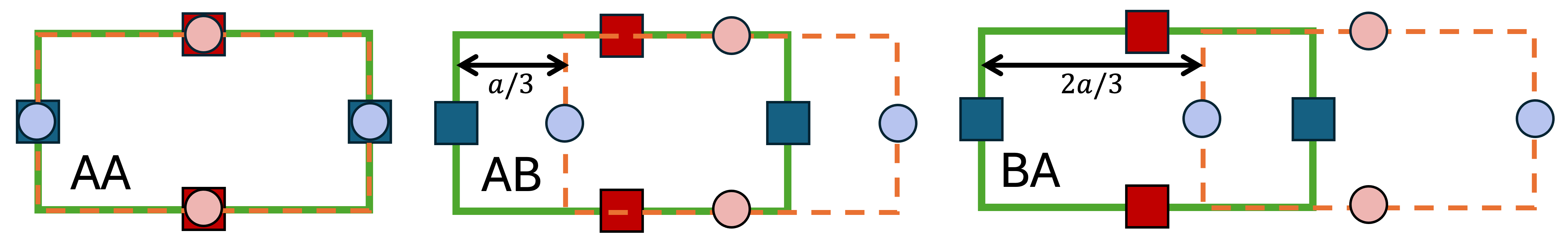}\\
\large{$+ \equiv \uparrow\downarrow / \uparrow\downarrow$, $- \equiv \uparrow\downarrow /\downarrow\uparrow$\\}
\includegraphics[width=0.45\textwidth]{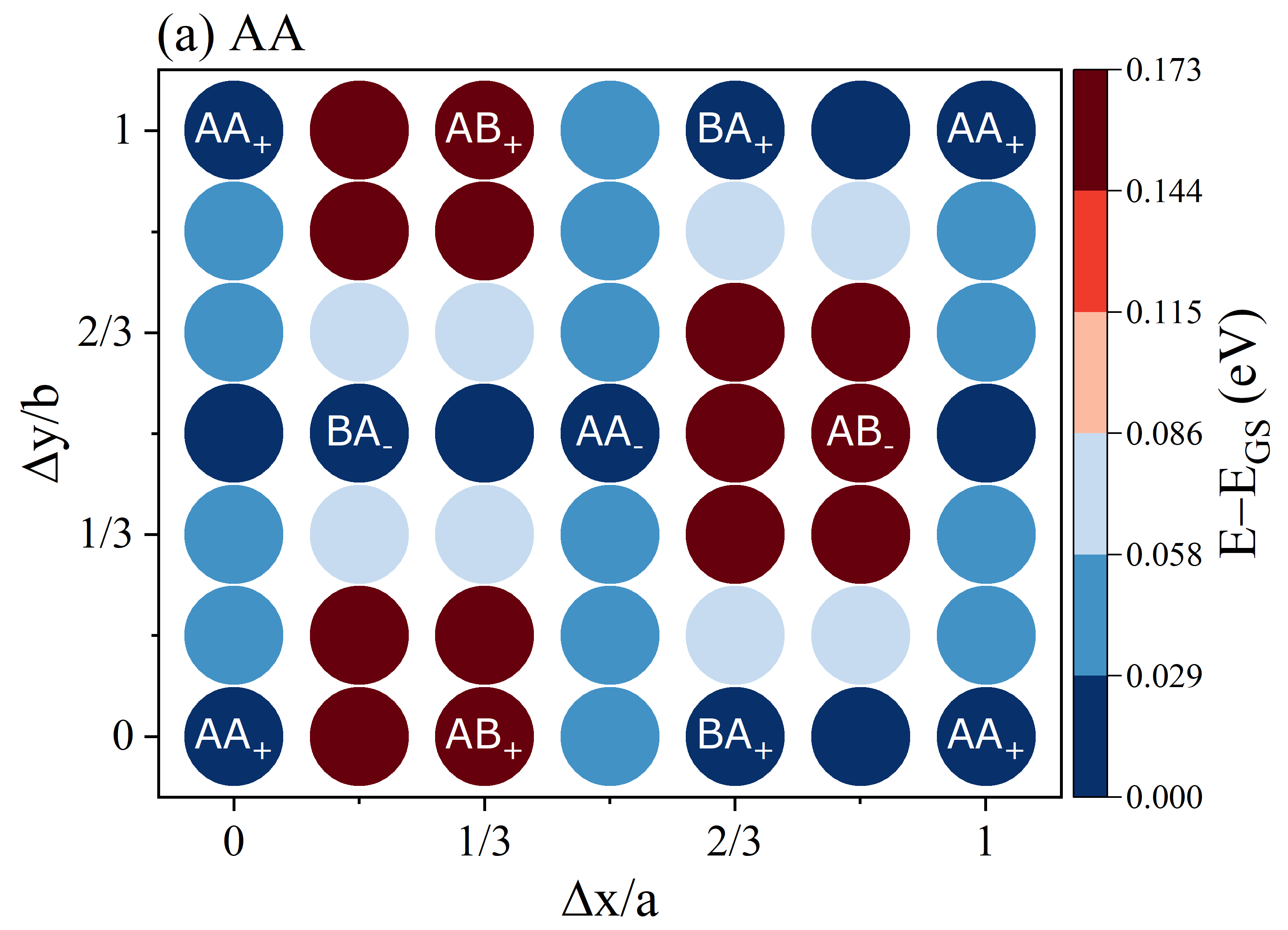}
\includegraphics[width=0.45\textwidth]{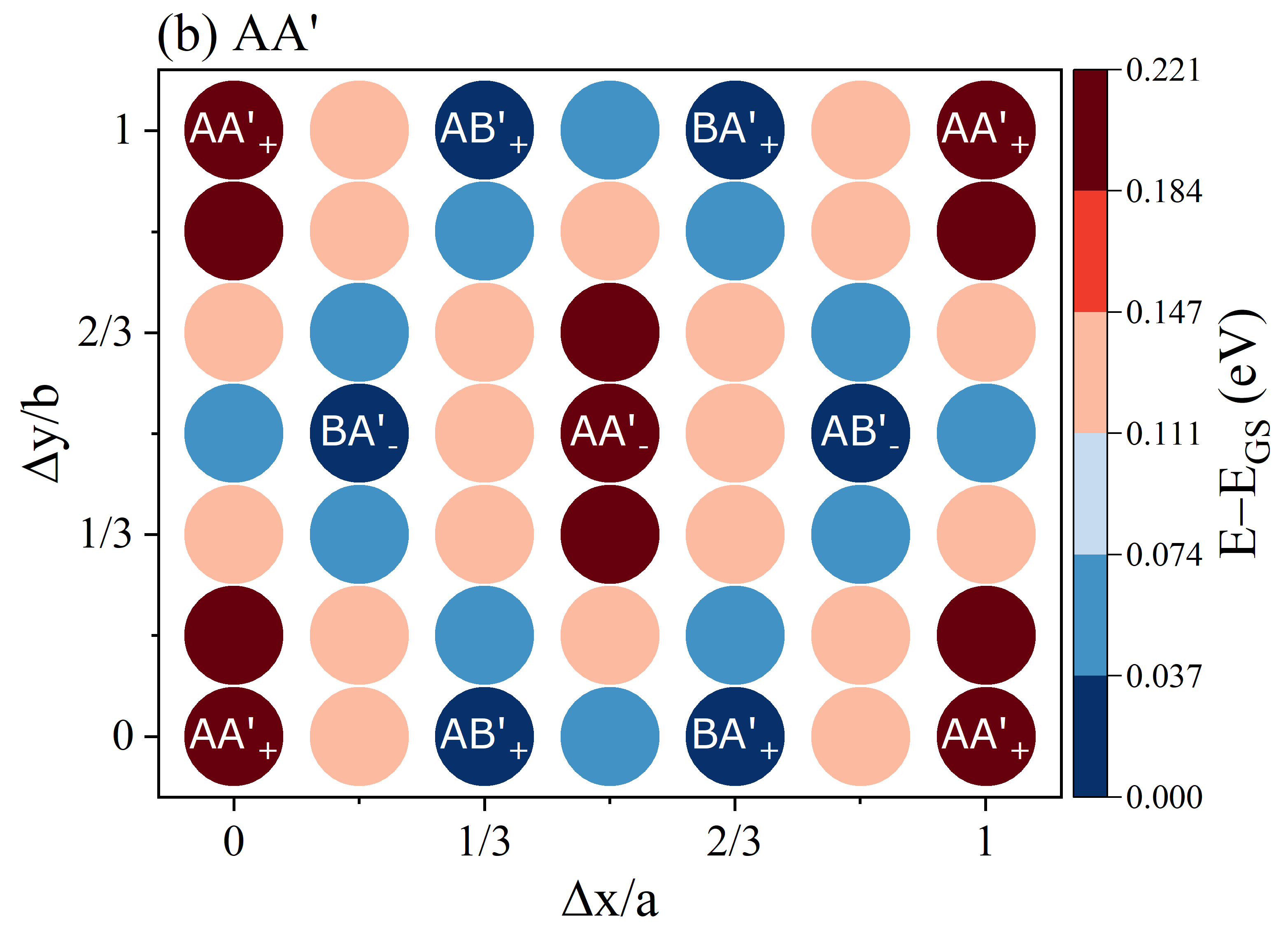}
\includegraphics[width=0.45\textwidth]{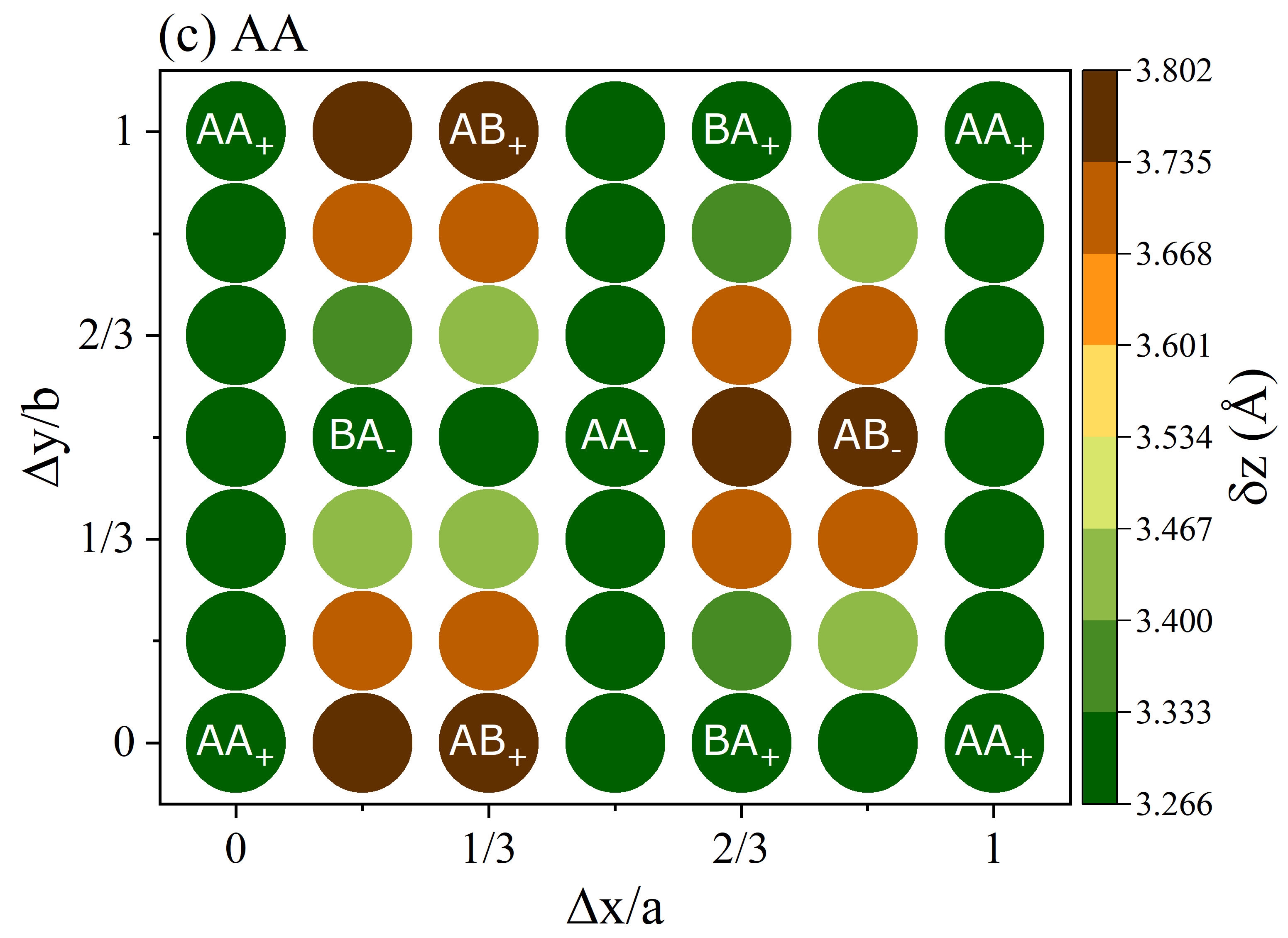}
\includegraphics[width=0.45\textwidth]{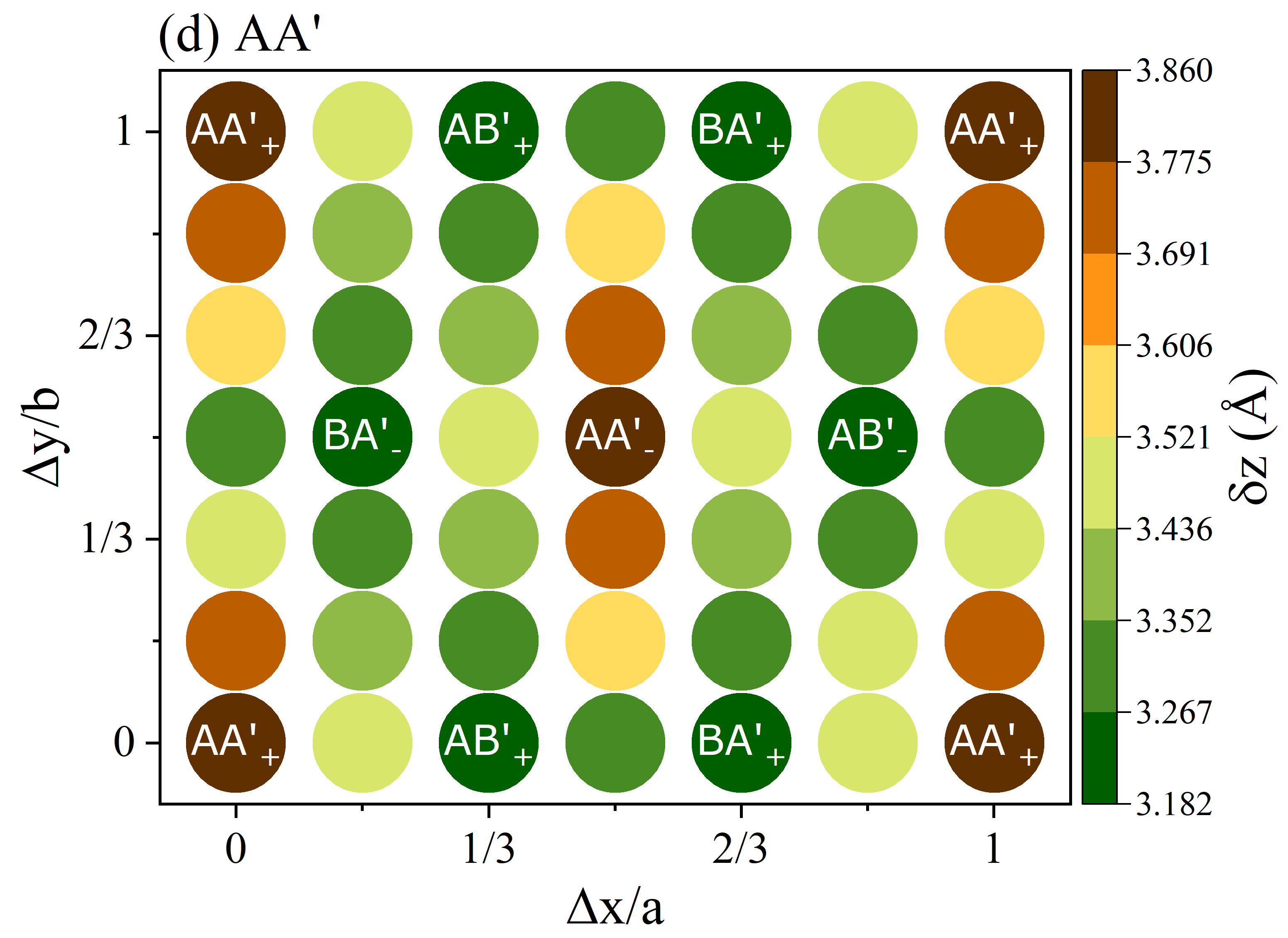}
\caption{Controlled in-plane displacement and magnetic configuration considering AFM monolayers. We use a subscript to label the magnetic state: “$+$” for the parallel $\uparrow\downarrow/\uparrow\downarrow$ and “$-$” for the antiparallel $\uparrow\downarrow/\downarrow\uparrow$ configurations. Starting from the AA$_{+}$ and AA$_{+}$($^{\prime}$) reference, the top schematics illustrate the highly symmetric registries AA($^{\prime}$), AB($^{\prime}$),  and BA($^{\prime}$). Panels \textbf{(a)}-\textbf{(b)} show the stacking relative energy $\Delta E=E-E_{\mathrm{GS}}$ as a function of the lateral shift $(\Delta x/a,\, \Delta y/b)$ for AA- and AA$^{\prime}$-based configurations, respectively. Panels \textbf{(c)}-\textbf{(d)} display the corresponding variation of the interlayer separation $\delta z$, defined as the distance between the inner iodine planes of opposite layers.}
\label{fig:2dneb} 
\end{figure*}

Our results show a consistent correlation between the most energetically stable bilayer configurations and those with the shortest interlayer distances (Table~\ref{tab:D3_per_atom}). Calculations using the DFT-D3(BJ) method yield average in-plane lattice parameters of 7.24~\AA{} and 3.89~\AA{}. The corresponding average interlayer distance is 3.42~\AA{}, although this value is notably sensitive to the specific stacking and magnetic order, as indicated by a standard deviation of 0.29~\AA{}. These interlayer separations are comparable to DFT-calculated values for CrI$_3$ bilayers ($3.3-3.5$ \AA)~\cite{leon2020strain} and slightly larger than theoretical predictions for CrBr$_3$ ($3.1-3.3$ \AA)~\cite{ghojavand2024strain}, reflecting the larger ionic radius of iodine compared to bromine and chlorine. Both structural parameters and energy hierarchies confirm that interlayer separation is highly dependent on the specific atomic registry and magnetic configuration, a general feature of van der Waals magnetic bilayers.

% D3BJ
\begin{table*}[]
\centering
\caption{Calculated properties of CrI$_2$ bilayers for direct (AA, AB, BA) and indirect (AA$^\prime$, AB$^\prime$, BA$^\prime$) stacking families using the DFT-D3(BJ) functional. For each stacking, we report the optimized in-plane lattice parameters ($|a|$, $|b|$), inner I--I vertical separation ($\delta_z$), and the lateral in-plane displacement between Cr atoms of opposite layers ($\delta_{xy}$). Energetic stability is shown by the relative energy within each stacking family ($\Delta E_{\text{local}}$) and relative to the global ground state ($\Delta E_{\text{Global}}$), which is the BA$^\prime$ configuration with intralayer AF$_x$ and interlayer antiparallel coupling. Distances are in \AA; energies are in meV per formula unit (f.u.).}
\label{tab:D3_per_atom}
\begin{tabular}{c|c|c|c|c|c|c|c}
\hline
 {Stacking} &
 {Magnetism} &
 {$|a|$} &
 {$|b|$} &
 {$\delta_z$} &
 {$\delta_{xy}$} &
 {$\Delta E_{\text{local}}$} &
 {$\Delta E_{\text{Global}}$} \\ \hline \hline
\multirow{4}{*}{AA}  & $\uparrow\downarrow/\uparrow\downarrow$  & 7.228 & 3.882 & 3.308 & 0.464 & \textbf{0.000} & 1.276  \\ \cline{2-8}
                     & $\uparrow\downarrow/\downarrow\uparrow$  & 7.245 & 3.884 & 3.322 & 0.434 & 0.099          & 1.375  \\ \cline{2-8}
                     & $\downarrow\downarrow/\uparrow\uparrow$  & 7.260 & 3.879 & 3.284 & 0.466 & 1.533          & 2.809  \\ \cline{2-8}
                     & $\uparrow\uparrow/\uparrow\uparrow$      & 7.300 & 3.891 & 3.205 & 0.509 & 0.910          & 2.185  \\ \hline
\multirow{4}{*}{AB}  & $\uparrow\downarrow/\uparrow\downarrow$  & 7.239 & 3.883 & 3.695 & 2.098 & 9.191          & 10.466 \\ \cline{2-8}
                     & $\uparrow\downarrow/\downarrow\uparrow$  & 7.239 & 3.883 & 3.695 & 2.098 & 8.926          & 10.202 \\ \cline{2-8}
                     & $\downarrow\downarrow/\uparrow\uparrow$  & 7.246 & 3.892 & 3.743 & 2.075 & 10.720         & 11.996 \\ \cline{2-8}
                     & $\uparrow\uparrow/\uparrow\uparrow$      & 7.256 & 3.896 & 3.234 & 2.271 & 1.555          & 2.831  \\ \hline
\multirow{4}{*}{BA}  & $\uparrow\downarrow/\uparrow\downarrow$  & 7.228 & 3.893 & 3.290 & 2.285 & 0.624          & 1.900  \\ \cline{2-8}
                     & $\uparrow\downarrow/\downarrow\uparrow$  & 7.213 & 3.896 & 3.269 & 2.276 & 0.330          & 1.605  \\ \cline{2-8}
                     & $\downarrow\downarrow/\uparrow\uparrow$  & 7.240 & 3.896 & 3.295 & 2.284 & 2.485          & 3.761  \\ \cline{2-8}
                     & $\uparrow\uparrow/\uparrow\uparrow$      & 7.246 & 3.899 & 3.291 & 2.283 & 1.675          & 2.950  \\ \hline \hline
\multirow{4}{*}{AA$^\prime$} & $\uparrow\downarrow/\uparrow\downarrow$ & 7.238 & 3.886 & 3.945 & 0.000 & 13.806         & 13.806 \\ \cline{2-8}
                     & $\uparrow\downarrow/\downarrow\uparrow$ & 7.243 & 3.885 & 3.937 & 0.000 & 13.612         & 13.612 \\ \cline{2-8}
                     & $\downarrow\downarrow/\uparrow\uparrow$ & 7.259 & 3.878 & 3.929 & 0.001 & 15.576         & 15.576 \\ \cline{2-8}
                     & $\uparrow\uparrow/\uparrow\uparrow$     & 7.258 & 3.880 & 3.932 & 0.000 & 15.678         & 15.678 \\ \hline
\multirow{4}{*}{AB$^\prime$} & $\uparrow\downarrow/\uparrow\downarrow$ & 7.237 & 3.888 & 3.216 & 2.161 & 0.485          & 0.485  \\ \cline{2-8}
                     & $\uparrow\downarrow/\downarrow\uparrow$ & 7.239 & 3.887 & 3.209 & 2.164 & 0.612          & 0.612  \\ \cline{2-8}
                     & $\downarrow\downarrow/\uparrow\uparrow$ & 7.268 & 3.893 & 3.237 & 2.163 & 2.531          & 2.531  \\ \cline{2-8}
                     & $\uparrow\uparrow/\uparrow\uparrow$     & 7.274 & 3.897 & 3.177 & 2.187 & 1.493          & 1.493  \\ \hline
\multirow{4}{*}{BA$^\prime$} & $\uparrow\downarrow/\uparrow\downarrow$ & 7.237 & 3.888 & 3.194 & 2.288 & 0.081          & 0.081  \\ \cline{2-8}
                     & $\uparrow\downarrow/\downarrow\uparrow$ & 7.237 & 3.888 & 3.194 & 2.288 & \textbf{0.000} & \textbf{0.000}  \\ \cline{2-8}
                     & $\downarrow\downarrow/\uparrow\uparrow$ & 7.257 & 3.893 & 3.205 & 2.293 & 2.165          & 2.165  \\ \cline{2-8}
                     & $\uparrow\uparrow/\uparrow\uparrow$     & 7.268 & 3.893 & 3.202 & 2.288 & 1.116          & 1.116  \\ \hline
\end{tabular}
\end{table*}

%% BANDAS
%% REVISAR
Upon transitioning from a monolayer to a bilayer system, the band structure in Figure \ref{fig:BL_bands} shows a reduction in the electronic band gap. In the antiferromagnetic ground state (AF$_x$), the CrI$_2$ monolayer exhibits a band gap of $E_g = 1.12$~eV (Fig.~\ref{fig:ML_bands}(left)), whereas, the bilayer configurations present slightly smaller values, ranging from 0.99 to 1.07 eV, depending on the stacking and interlayer magnetic alignment. This reduction primarily arises from interlayer orbital hybridization, where the out-of-plane I-$p_z$ orbitals and Cr-$d$ orbitals couple across the van der Waals gap, forming bonding and antibonding states that enhance band dispersion. This interaction typically raises the valence band maximum and lowers the conduction band minimum, thereby narrowing the band gap.  
The magnitude of the gap does not follow a simple trend with interlayer distance $\delta z$, reflecting the joint role of registry, orbital character, and interlayer alignment. For example, the BA$^\prime$ stacking has the shortest interlayer separation ($\delta z \sim 3.19-3.20$~\AA, Table~\ref{tab:D3_per_atom}), yet its gap ($E_g \sim 1.03-1.04$~eV) is larger than that of the AA$({\uparrow\downarrow/\downarrow\uparrow}$) configuration ($E_g = 0.99$~eV), which has a longer separation ($\delta z \sim 3.31$~\AA). This non-monotonic behavior indicates that the gap is not determined solely by interlayer proximity, but rather by a subtle interplay of stacking symmetry, orbital character, and magnetic order.  

In the BA$^\prime$ configuration (Fig.~\ref{fig:BL_bands}(right)), the broken inversion symmetry and absence of high-symmetry axes lift band degeneracies through avoided crossings, hindering the gap-reducing effect of strong interlayer coupling. Moreover, interlayer hybridization is orbital-selective: while certain overlaps (e.g., I-$p_z$/Cr-$d_{z^2}$) reduce the gap by enhancing dispersion, others redistribute spectral weight without closing the fundamental gap. Finally, the interlayer magnetic alignment directly modulates the exchange splitting at the band edges. This effect not only alters the overall band gap but also lifts the spin degeneracy of the electronic states.
The lifting of spin degeneracy, which manifests as small spin splittings in the band structure, is a key feature of the BA$^\prime$ and other indirect bilayers. 
\bluemark{The resulting band splittings are relatively small and become clearly visible only upon magnification of the valence-band region (shown in Fig.~S7).}
The effect originates from the exchange interaction and is present in our calculations even without spin-orbit coupling, arising from the simultaneous breaking of time-reversal symmetry ($\mathcal{T}$) by the antiferromagnetic order and inversion symmetry ($\mathcal{P}$) by the non-centrosymmetric stacking. While the absence of the combined $\mathcal{PT}$ symmetry is the prerequisite, the splitting itself results from the effective exchange fields imposed by the magnetic order within a structurally polar environment. In this sense, broken inversion symmetry enables asymmetric interlayer exchange coupling and sublattice-dependent potential gradients, which, together with the intrinsic AF exchange, give rise to finite and momentum-dependent spin splittings. Although spin-orbit coupling may introduce additional relativistic corrections, such as Dzyaloshinskii-Moriya interactions, the spin splitting discussed here is primarily driven by exchange interactions in a non-centrosymmetric crystal structure.
%\bluemark{In this context, it is important to emphasize that the present CrI$_2$ bilayers do not satisfy the symmetry requirements for altermagnetism. According to the general symmetry classification of altermagnets, altermagnetic band splitting requires specific non-relativistic spin-group symmetries that connect opposite-spin sublattices via crystallographic rotations or reflections without invoking time-reversal symmetry. Such symmetry operations are absent in all stacking configurations considered here, as established within the general framework developed in Ref.~\cite{gonzalez2025engineering,pan2024general}. Therefore, the exchange-induced splittings observed in CrI$_2$ bilayers should not be interpreted as signatures of altermagnetic order.}

The magnitude of these splittings is highly sensitive to the interlayer magnetic alignment. In the parallel $\uparrow\downarrow/\uparrow\downarrow$ configuration, the exchange fields from the two layers add constructively, leading to pronounced band-edge splittings, 8.3~meV (VBM) and 11.6~meV (CBM) at the $\Gamma$ point, and similarly large values of 8.8~meV (VBM) and 12.1~meV (CBM) at the Y point. By contrast, in the antiparallel $\uparrow\downarrow/\downarrow\uparrow$ configuration, although the stacking likewise breaks $\mathcal{PT}$ symmetry, the opposing exchange fields largely cancel. This cancellation yields much smaller splittings, with the VBM and CBM reaching only 5.0~meV and 3.2~meV at $\Gamma$, and reducing further to 4.2~meV and 0.7~meV at M.

\bluemark{The CrI$_2$ bilayers considered here do not satisfy the symmetry requirements for altermagnetism. According to the general classification~\cite{zeng2024description,gonzalez2025engineering,pan2024general}, altermagnetic band splitting requires non-relativistic spin-group symmetries that connect opposite-spin sublattices through crystallographic rotations or reflections without invoking time-reversal symmetry. In the present case, the structurally stable and metastable bilayer stackings investigated in this work realize the space groups $Pm$ (No.~6) for BA$'$, $Cm$ (No.~8) for AB$'$, and $Amm2$ (No.~38) for AA$'$, while the direct stackings (BA, AB, and AA) belong to the centrosymmetric space group $C2/m$ (No.~12). As none of these space groups contain the requisite spin-group symmetries, the symmetry conditions for altermagnetism are not fulfilled. In particular, they lack rotation or mirror operations that map opposite-spin sublattices onto each other. Therefore, the exchange-induced band splittings observed in CrI$_2$ bilayers should not be interpreted as signatures of altermagnetic order.}

\subsection{Bilayer: Controlled sliding}

To gain insight into the relationship between stacking and magnetism, we map the potential energy surface by performing controlled in-plane displacement of one monolayer relative to the other, as illustrated in Fig.~\ref{fig:2dneb}. These calculations use the DFT-D3(BJ) functional with lattice constants from the relaxed ground state. For both AA and AA$^\prime$ families, we systematically map the potential energy surface with full electronic and ionic relaxation at each step. The outermost iodine positions are constrained to maintain registry while all other atoms freely relax, with magnetic moments on the four Cr atoms fixed.

We introduce a simplified subscript notation to identify magnetic states: '$+$' represents parallel interlayer alignment ($\uparrow\downarrow/\uparrow\downarrow$) and '$-$' denotes antiparallel alignment ($\uparrow\downarrow/\downarrow\uparrow$). This systematic approach connects all high-symmetry stackings through fractional translations, with registries interconnected by specific displacement vectors: $AA_-=AA_++(1/2,1/2)$, $AB_+=AA_++(1/3,0)$, $AB_-=AA_++(5/6,1/2)$, $BA_+=AA_++(2/3,0)$, and $BA_-=AA_++(1/6,1/2)$.

The corrugated energy surface reveals distinct stable minima corresponding to different stacking registries, separated by moderate barriers of $\sim 25-50$ meV/f.u, establishing direct pathways for stacking engineering, where mechanical sliding can switch between stable states~\cite{gonzalez2025engineering}. Because each registry carries different point-group symmetry, sliding toggles functionalities: breaking inversion symmetry activates interlayer ferroelectricity and Dzyaloshinskii-Moriya interactions, providing handles to control polarity and chiral exchange~\cite{acharya2025first}.

The interlayer separation $\delta z$ follows the registry, growing when inner I atoms face each other and shrinking when lateral offsets reduce steric overlap. BA$^\prime$ minimizes I--I repulsion and hosts the global minimum, consistent with our relaxed energetics and stacking-driven symmetry breaking~\cite{acharya2025first}, complementing strain-based magnetic tunability~\cite{widyandaru2025tunable}. This tunability enables advanced spintronic applications: DMI control via stacking can host chiral magnetic textures and skyrmions for racetrack concepts~\cite{acharya2025first,parkin2015memory,fert2013skyrmions}, while low energy barriers enable low-power straintronics and magnonic applications~\cite{gonzalez2025engineering,widyandaru2025tunable,chumak2015magnon}.

%% NUEVO SOBRE J
\subsection{Exchange parameters}

After a structural relaxation and self-consistent calculation within \textsc{OpenMX}, the resulting Hamiltonians are processed with \textsc{TB2J} to obtain pairwise exchange tensors $\mathsf J_{ij}$ via the magnetic force theorem.
A custom post-processing script groups crystallographically equivalent bonds into distance shells and separates in-plane from interlayer pairs using small geometric tolerances. For each shell, we report: (i) the representative Cr-Cr distance \(d\), (ii) the number of distinct bonds \(n\) in the simulation cell, and (iii) shell-averaged couplings. In the no SOC tables, we list \(J_{\mathrm{eff}}\), defined as the shell average of \(J_{ij}\). In the SOC tables we list \(J^{\mathrm{iso}}\) and the effective directional couplings
\begin{equation}
J_\alpha = J^{\mathrm{iso}} + \langle \Gamma_{\alpha\alpha}\rangle \qquad (\alpha=x,y,z),
\end{equation}
that is, \(J_x\), \(J_y\), and \(J_z\). We quantify the in-plane symmetric exchange anisotropy as
\begin{equation}
\Delta J_{xy} \equiv J_x - J_y = \langle \Gamma_{xx}-\Gamma_{yy}\rangle .
\end{equation}
The Dzyaloshinskii-Moriya entry reports the shell average of the magnitude \(\langle |\mathbf D| \rangle\) and, when relevant, the magnitude of the vector average \(|\langle \mathbf D \rangle|\). When a line is labeled “both layers”, \(n\) equals the sum of symmetry equivalent bonds from the bottom and top layers (for example, \(n=8\) corresponds to \(4+4\)). All averages are taken over unique bonds after canonicalizing pair orientations. Interlayer entries include only pairs that connect different layers. Uncertainties from distance clustering remain below the last digit reported and do not affect comparative trends across stackings.

\begin{table*}[!t]
\centering
\caption{Nearest neighbour in plane exchange couplings for A-B bonds, $J^{AB}_{\text{in plane}}$, and for A-A/B-B bonds, $J^{\mathrm{AA/BB}}_{\text{in plane}}$, in CrI$_2$ monolayer and bilayers. Columns list the Cr-Cr distance $d$ (Å), number of bonds $n$, the scalar (no SOC) coupling $J_{\mathrm{eff}}$ (meV), the isotropic (SOC) part $J^{\mathrm{iso}}$ (meV), the shell-averaged DMI magnitude $\langle|\mathbf D|\rangle$ (meV), and the in plane symmetric exchange anisotropy $\Delta J_{xy}=J_x-J_y$ (meV), with $J_\alpha=J^{\mathrm{iso}}+\langle J^{\mathrm{ani}}_{\alpha\alpha}\rangle$. Positive $J$ denotes ferromagnetic coupling in $H=-\sum_{ij}\mathbf S_i^{\mathsf t}\mathsf J_{ij}\mathbf S_j$. The "Magnetism" column indicates the interlayer textures $\uparrow\downarrow/\uparrow\downarrow$ and $\uparrow\downarrow/\downarrow\uparrow$.}
\label{tab:J_inplane}
{Nearest-neighbour  in-plane A-B exchange couplings $J^{AB}_{\text{in-plane}}$\\[2pt]}
\begin{tabular}{l | c| c| c |c |c| c |c}
\hline
System & Magnetism & $d$ (Å) & $n$ & $J^{AB}_{\mathrm{eff}}$ (no SOC) & $J^{AB}_{\mathrm{iso}}$ (SOC) & $\langle|{\bf D}|\rangle$ & $\Delta J_{xy}$ \\
\hline\hline
\multirow{2}{*}{AA}        & $\uparrow\downarrow/\uparrow\downarrow$ & 4.10 & 8 & -4.01 & -4.84 & 0.02 & 1.49 \\
                           & $\uparrow\downarrow/\downarrow\uparrow$ & 4.10 & 8 & -4.00 & -4.83 & 0.01 & 1.48 \\ \hline
\multirow{2}{*}{AA$^\prime$} & $\uparrow\downarrow/\uparrow\downarrow$ & 4.11 & 8 & -3.85 & -4.70 & 0.01 & 1.48 \\
                           & $\uparrow\downarrow/\downarrow\uparrow$ & 4.11 & 8 & -3.87 & -4.72 & 0.02 & 1.49 \\ \hline
\multirow{2}{*}{AB$^\prime$} & $\uparrow\downarrow/\uparrow\downarrow$ & 4.10 & 8 & -3.88 & -4.78 & 0.02 & 1.48 \\
                           & $\uparrow\downarrow/\downarrow\uparrow$ & 4.10 & 8 & -3.99 & -4.82 & 0.01 & 1.47 \\ \hline
\multirow{2}{*}{BA$^\prime$} & $\uparrow\downarrow/\uparrow\downarrow$ & 4.11 & 8 & -4.00 & -4.83 & 0.01 & 1.47 \\ 
                           & $\uparrow\downarrow/\downarrow\uparrow$ & 4.11 & 8 & -4.01 & -4.83 & 0.02 & 1.47 \\  \hline
Monolayer & AF$_x$ $\uparrow\downarrow$                  & 4.11 & 4 & -3.66 & -4.50 & 0.00 & 1.48 \\
\hline
\end{tabular}
\vspace{6pt}

{Nearest-neighbour in-plane A-A / B-B exchange couplings
$J^{\mathrm{AA/BB}}_{\text{in-plane}}$}\\[2pt]
\begin{tabular}{l |c| c| c |c| c| c| c}
\hline
System & Magnetism & $d$ (Å) & $n$ & $J_{\mathrm{eff}}$ (no SOC) & $J_{\mathrm{iso}}$ (SOC) & $\langle|{\bf D}|\rangle$ & $\Delta J_{xy}$ \\
\hline\hline
\multirow{2}{*}{AA}         & $\uparrow\downarrow/\uparrow\downarrow$ & 3.89 & 8 & 2.79 & 2.86 & 0.01 & 1.00 \\
                            & $\uparrow\downarrow/\downarrow\uparrow$ & 3.89 & 8 & 2.77 & 2.85 & 0.02 & 0.94 \\ \hline
\multirow{2}{*}{AA$^\prime$}& $\uparrow\downarrow/\uparrow\downarrow$ & 3.89 & 8 & 2.73 & 2.80 & 0.01 & 1.00 \\
                            & $\uparrow\downarrow/\downarrow\uparrow$ & 3.89 & 8 & 2.74 & 2.82 & 0.01 & 0.98 \\ \hline
\multirow{2}{*}{AB$^\prime$}& $\uparrow\downarrow/\uparrow\downarrow$ & 3.89 & 8 & 2.65 & 2.74 & 0.02 & 1.00 \\
                            & $\uparrow\downarrow/\downarrow\uparrow$ & 3.89 & 8 & 2.70 & 2.78 & 0.03 & 1.01 \\ \hline
\multirow{2}{*}{BA$^\prime$}& $\uparrow\downarrow/\uparrow\downarrow$ & 3.89 & 8 & 2.72 & 2.79 & 0.02 & 1.01 \\
                            & $\uparrow\downarrow/\downarrow\uparrow$ & 3.89 & 8 & 2.73 & 2.81 & 0.02 & 1.06 \\ \hline
Monolayer & AF$_x$ $\uparrow\downarrow$                                & 3.89 & 4 & 2.56 & 2.64 & 0.00 & 1.03 \\
\hline
\end{tabular}
\end{table*}

Table~\ref{tab:J_inplane} summarizes the leading in-plane exchange channels in CrI$_2$: the inter-chain antiferromagnetic coupling $J^{AB}_{\text{in\,plane}}$ and the intra-chain ferromagnetic coupling $J^{\mathrm{AA/BB}}_{\text{in\,plane}}$. We use angle brackets $\langle\cdot\rangle$ to denote shell averages within a given registry, and $\langle\cdot\rangle_{\mathrm{reg}}$ denotes an average taken over all bilayer registries. 
Positive $J$ favors ferromagnetic alignment, negative $J$ favors antiferromagnetic alignment.
The layer count sets the primary magnetic energy scale. Relative to the monolayer reference ($J^{AB}_{\mathrm{eff}}=-3.66$\,meV and $J^{AB}_{\mathrm{iso}}=-4.50$\,meV), every bilayer strengthens the nearest-neighbour A-B exchange. Averaging over all registries gives $\langle J^{AB}_{\mathrm{eff}}\rangle_{\mathrm{reg}}=-3.95\pm0.06$\,meV and $\langle J^{AB}_{\mathrm{iso}}\rangle_{\mathrm{reg}}=-4.79\pm0.05$\,meV, an increase of about 8 to 10\% in magnitude. The spread across registries remains modest ($\sim 0.16$\,meV). The interlayer magnetic alignment produces much smaller splits at fixed registry (typically $\leq 0.01$\,meV for AA and BA$^\prime$, about $0.01$\,meV for AA$^\prime$, and reaching $\sim 0.11$\,meV only in AB$^\prime$). The intra-chain channel follows the same pattern: $J^{\mathrm{AA/BB}}_{\mathrm{eff}}$ increases from 2.56\,meV in the monolayer to $\langle J^{\mathrm{AA/BB}}_{\mathrm{eff}}\rangle_{\mathrm{reg}}=2.72\pm0.04$\,meV in bilayers (about 6\%). The nearly unchanged Cr-Cr distance across registries ($d\sim 4.10$ to $4.11$\,\AA) points to stacking-controlled orbital hybridization pathways rather than bond-length effects as the driver of the energy scaling, consistent with broader trends reported for stacking engineering in van der Waals bilayers \cite{chen2025tunable,ji2023general}.

Symmetry considerations explain the anisotropy data. Inversion symmetry forbids the Dzyaloshinskii-Moriya interaction, so the centrosymmetric monolayer exhibits exactly zero DMI by symmetry, while non-centrosymmetric bilayer stackings enable finite DMI values in the 0.01 to 0.02\,meV range. In contrast, the symmetric exchange anisotropy shows internal robustness: $\Delta J_{xy}$ clusters near 1.48\,meV for A-B bonds and near 1.03\,meV for A-A/B-B bonds, with weak dependence on layer count and registry. The pattern supports a local bond-environment origin for the symmetric anisotropy and a symmetry-enabled, bilayer-level origin for the DMI \cite{chen2025tunable,ji2023general}.
The combined behaviour of $J^{AB}_{\text{in\,plane}}$ and $J^{\mathrm{AA/BB}}_{\text{in\,plane}}$ yields a coherent microscopic picture for the stripe antiferromagnetic ground state. Stacking controls the absolute scale of both competing channels and tunes the energy balance that selects the ordered phase. Interlayer magnetic alignment provides a secondary knob for fine adjustments. From a device perspective, in-plane sliding sets the stacking registry and therefore the relevant magnetic energy scale, while symmetry selection rules govern which anisotropic and chiral interaction terms are allowed.

\begin{table}[!t]
\centering
\caption{Nearest-neighbour  interlayer exchange couplings for CrI$_2$ bilayers. The table follows the same distribution of Table \ref{tab:J_inplane}.}
\label{tab:J_interlayer}
% ---------- Shell 1 ----------
%{Nearest-neighbour interlayer exchange coupling }\\[2pt]
\begin{tabular}{l |c |c| c| c| c| c| c}
\hline
Syst. & Mag. & $d$ (Å) & $n$ & $J_{\mathrm{eff}}$  & $J^{\mathrm{iso}}$  & $\langle|{\bf D}|\rangle$ & $\Delta J_{xy}$ \\
  &   &  &  &  (no SOC) &  (SOC) &  &  \\
\hline\hline
\multirow{2}{*}{AA}         & $\uparrow\downarrow/\uparrow\downarrow$ & 6.94 & 2 &  \,0.04 &  \,0.05 & 0.00 &  \,0.16 \\
                            & $\uparrow\downarrow/\downarrow\uparrow$ & 6.94 & 2 & -0.05 & -0.07 & 0.00 & -0.02 \\ \hline
\multirow{2}{*}{AA$^\prime$}& $\uparrow\downarrow/\uparrow\downarrow$ & 7.55 & 2 &  \,0.11 & -0.10 & 0.08 &  \,0.36 \\
                            & $\uparrow\downarrow/\downarrow\uparrow$ & 7.55 & 2 & -0.20 &  \,0.16 & 0.08 & -0.03 \\ \hline
\multirow{2}{*}{AB$^\prime$}& $\uparrow\downarrow/\uparrow\downarrow$ & 7.23 & 6 & -0.05 & -0.07 & 0.02 &  \,0.08 \\
                            & $\uparrow\downarrow/\downarrow\uparrow$ & 7.25 & 6 &  \,0.04 &  \,0.04 & 0.02 &  \,0.27 \\ \hline
\multirow{2}{*}{BA$^\prime$}& $\uparrow\downarrow/\uparrow\downarrow$ & 7.26 & 6 &  \,0.04 &  \,0.03 & 0.02 &  \,0.26 \\
                            & $\uparrow\downarrow/\downarrow\uparrow$ & 7.26 & 6 & -0.05 & -0.07 & 0.02 &  \,0.08 \\
\hline
\end{tabular}

%\vspace{6pt}

% ---------- Shell 2 ----------
%{Second-neighbour interlayer exchange coupling }\\[2pt]
%\begin{tabular}{l c c c c c c c}
%\hline
%System & Magnetism & $d$ (Å) & $n$ & $J_{\mathrm{eff}}$ (no SOC) & $J^{\mathrm{iso}}$ (SOC) & $\langle|{\bf D}|\rangle$ & $\Delta J_{xy}$ \\
%\hline\hline
%\multirow{2}{*}{AA}         & $\uparrow\downarrow/\uparrow\downarrow$ & 7.77 & 4 & -0.09 & -0.12 & 0.00 & -0.20 \\
 %                           & $\uparrow\downarrow/\downarrow\uparrow$ & 7.77 & 4 &  \,0.08 &  \,0.09 & 0.00 &  \,0.69 \\
%\multirow{2}{*}{AA$^\prime$}& $\uparrow\downarrow/\uparrow\downarrow$ & 8.49 & 6 &  \,0.01 &  \,0.01 & 0.00 &  \,0.27 \\
 %                           & $\uparrow\downarrow/\downarrow\uparrow$ & 8.50 & 4 & -0.02 & -0.03 & 0.00 & -0.14 \\
%\multirow{2}{*}{AB$^\prime$}& $\uparrow\downarrow/\uparrow\downarrow$ & 8.26 & 4 &  \,0.19 &  \,0.19 & 0.01 &  \,0.40 \\
%                            & $\uparrow\downarrow/\downarrow\uparrow$ & 8.26 & 4 & -0.17 & -0.17 & 0.00 & -0.11 \\
%\multirow{2}{*}{BA$^\prime$}& $\uparrow\downarrow/\uparrow\downarrow$ & 8.26 & 4 & -0.17 & -0.18 & 0.00 & -0.11 \\
%                            & $\uparrow\downarrow/\downarrow\uparrow$ & 8.26 & 4 &  \,0.19 &  \,0.19 & 0.01 &  \,0.41 \\
%\hline
%\end{tabular}
\end{table}

Interlayer exchange remains one order of magnitude weaker than the in-plane channels, yet it displays clear and symmetry-driven trends. The nearest-neighbour shell in Table~\ref{tab:J_interlayer} shows a sharp contrast between direct and indirect stacking. Direct AA stacking keeps an almost vanishing Dzyaloshinskii-Moriya magnitude, $\langle|\mathbf D|\rangle\!=\!0$, and the smallest in-plane symmetric anisotropy, $|\Delta J_{xy}|\!\lesssim\!0.16$~meV. Indirect registries AA$^\prime$, AB$^\prime$, and BA$^\prime$ break inversion and develop finite $\langle|\mathbf D|\rangle$ in the $0.02$-$0.08$~meV range together with larger $\Delta J_{xy}$ values, up to $0.36$~meV in AA$^\prime$. This behaviour agrees with the stacking-controlled symmetry breaking reported for van der Waals bilayers \cite{chen2025tunable,ji2023general}.
The Heisenberg part is small in all cases, with typical $|J^{\mathrm{iso}}|\!\sim\!0.03$--$0.07$~meV, and it often changes sign when the interlayer magnetic alignment switches between $\uparrow\downarrow/\uparrow\downarrow$ and $\uparrow\downarrow/\downarrow\uparrow$. AA flips from $-0.07$ to $+0.05$~meV, AA$^\prime$ from $+0.16$ to $-0.10$~meV, AB$^\prime$ from $+0.04$ to $-0.07$~meV, and BA$^\prime$ from $-0.07$ to $+0.03$~meV. These reversals indicate a weak and highly tunable interlayer exchange, while the anisotropic terms follow the stacking symmetry more rigidly.

Second-neighbour interlayer couplings remain below $0.2$~meV across all registries and do not alter the qualitative picture established by the nearest-neighbour shell. Most anisotropy values are modest; the main outlier is AA with $\uparrow\downarrow/\downarrow\uparrow$, where $\Delta J_{xy}\!\sim\!0.69$~meV appears alongside a small $J^{\mathrm{iso}}\!\sim\!0.09$~meV. Given their small magnitude relative to the leading in-plane exchanges, the second-neighbour numbers are not shown.

The exchange mapping yields a coherent microscopic picture and establishes a clear ordering of interactions. The two nearest-neighbor in-plane channels, the inter-chain antiferromagnetic coupling $J^{AB}_{\text{in\,plane}}$ and the intra-chain ferromagnetic coupling $J^{\mathrm{AA/BB}}_{\text{in\,plane}}$, set the magnetic energy scale. Across all registries, bilayer formation strengthens both channels by approximately 6-10\% relative to the monolayer, while the symmetric in-plane anisotropy remains nearly registry-independent: $\Delta J_{xy}\sim 1.48$~meV for A-B bonds and $\sim 1.03$~meV for A-A/B-B bonds. Dzyaloshinskii-Moriya interactions emerge exclusively when stacking breaks inversion symmetry, consistent with recent findings in stacking-engineered van der Waals bilayers~\cite{chen2025tunable,ji2023general,gui2025antiparallel}. Interlayer exchange remains an order of magnitude weaker; however, its sign depends sensitively on both registry and interlayer magnetic alignment, enabling fine-tuning of canting angles and spin-flop fields without disrupting the primary in-plane energy scale. This comprehensive parameter set provides a compact, transferable framework for subsequent finite-temperature and spin-dynamics analyses.

%%%%%%%%
%%%%%%%%

\subsection{Electronic Origin of Magnetic Exchange}
To understand the trends in magnetic couplings (in Tables \ref{tab:J_inplane} and \ref{tab:J_interlayer}) from an electronic point of view, we evaluate spin-resolved band centers for the Cr-$d$ and I-$p$ manifolds from the projected density of states \(\rho\). The band center of a given projection \(X\!\in\!\{p,d\}\) and spin \(\sigma\) is defined as the first moment of the $\rho$,
\begin{equation}
\varepsilon^{\sigma}_X(X)=
\frac{\displaystyle\int  E\,\rho_X^{\sigma}(E)\,dE}
{\displaystyle\int \rho_X^{\sigma}(E)\,dE},
\end{equation}
with energies measured relative to the Fermi level of each calculation. We include 100 Kohn-Sham bands per spin (including valence and conduction) to ensure spectral convergence of the moments.

For the AFM monolayer, the centre of the majority-spin Cr-$d^\uparrow$ band is $-2.42$ eV and the centre of the I-$p$ band is $-2.56$ eV, giving a small energy mismatch of $\Delta \varepsilon_{pd}^\uparrow = 0.16$ eV. 
Whereas, for the minority-spin Cr-$d^\downarrow$ states, the band center appears at a higher energy, 3.32 eV, exhibiting a significant energy mismatch $\Delta \varepsilon_{pd}^\downarrow = 5.94$ eV. 
For the bilayer ground state, the BA$^\prime$ exhibits a similar order in the band centers, with $-2.44$ eV for the Cr-$d^\uparrow$ states and 3.35 eV for the Cr-$d^\downarrow$ states. For I-$p$ band centers, we consider an inner I-$p$ band center ($-2.66$ eV) and an outer one ($-2.60$ eV), corresponding to a slight difference ($ \sim 0.06$ eV) between the I-$p$ band centers. The difference in the band centers yields $\Delta \varepsilon_{pd}^\uparrow \sim 0.16-0.19$ eV and $\Delta \varepsilon_{pd}^\downarrow \sim 6.0$ eV. While the difference between inner-outer I-$p$ centers does not affect the fundamental mechanism, it signals the onset of symmetry breaking in the bilayer,  a prerequisite for anisotropic interactions.

The distribution of the p/d band centers suggest a highly spin-selective superexchange pathway, in which only the Cr-$d$ majority spin states hybridize effectively with the $p$ orbitals of iodine, supporting the robust antiferromagnetic order observed in both monolayer and bilayer structures. Given that the distribution of band centers remains quantitatively the same across configurations, we can attribute it to an intrinsic behavior of CrI$_2$, which we associate with the Anderson-Goodenough-Kanamori picture for high-spin Cr$^{2+}$ ($d^4$)~\cite{goodenough1955theory,kanamori1959superexchange,anderson1959new}, providing an electronic explanation for the observed magnetic scheme, where $J$ is controlled both by the energy mismatch and, more sensitively, by the transfer integral $t$, itself highly responsive to bond angles, interlayer separations, and local crystal fields.

\begin{table}[t]
\centering
\caption{In-plane polarization $P_x$ relative to BA$^\prime$ registry 
%($P_x(\mathrm{BA}')=0$). 
A positive $\Delta P_x$ denotes a dipole pointing along $+x$ (as defined in Fig.~\ref{fig:scheme}), while a negative value points along $-x$. By symmetry, reversing the stacking registry or interlayer magnetic alignment flips the sign. 
%Reported values are Berry-phase differences "unfolded" to the minimum-norm branch. 
$\Delta P_x$ is given in $\mu$C/cm$^2$, and the corresponding dipole-per-area is $\Delta(p/A)=\Delta P_x \cdot c$ (with $c=24$~\AA) in nC/m. 
}
\label{tab:pol}
\begin{tabular}{l|c|c |c|c|c }
\hline
\multirow{2}{*}{Configuration} & \multirow{2}{*}{SOC} & \multicolumn{2}{c|}{AA$^\prime$-BA$^\prime$} & \multicolumn{2}{c}{AB$^\prime$-BA$^\prime$} \\
              & &$\Delta P_x$ & $\Delta(p/A)$ & $\Delta P_x$ & $\Delta(p/A)$ \\
\hline \hline
\multirow{2}{*}{$\uparrow\downarrow/\downarrow\uparrow$}& no SOC & $-8.52$ & $-0.20$ & $+7.66$ & $+0.18$ \\
                                                        &   SOC    & $+0.35$ & $+0.01$ & $-0.67$ & $-0.02$ \\ \hline
\multirow{2}{*}{$\uparrow\downarrow/\uparrow\downarrow$} &no SOC & $+1.50$ & $+0.04$ & $+1.17$ & $+0.03$ \\
                                                        & SOC   & $+1.48$ & $+0.04$ & $+1.12$ & $+0.03$ \\
\hline
\end{tabular}
\end{table}

\subsection{Magnetoelectric Effects from Stacking-Induced Symmetry Breaking}

Indirect bilayer stackings such as AA$^\prime$, BA$^\prime$, and AB$^\prime$ break inversion symmetry, activating both finite Dzyaloshinskii-Moriya interactions and magnetoelectric coupling. In BA$^\prime$, for example, the small ($\sim$0.05-0.06 eV) splitting between inner and outer iodine $p$ levels reflects ligand inequivalence, a microscopic signature of inversion breaking. Symmetry analysis shows that while the centrosymmetric AFM monolayer forbids any polarization, BA$^\prime$ retains a mirror plane along $y$, which constrains a spontaneous polarization $P_x$ along the in-plane $x$ direction. The same symmetry reduction also permits finite DMI vectors and registry-dependent symmetric anisotropy.

While recent work on the orthorhombic phase of CrI$_2$ has highlighted sliding ferroelectricity with out-of-plane polarization driven by exchange-striction~\cite{yu2025sliding}, our results reveal a different mechanism in monoclinic bilayers, where stacking registry enables in-plane polarization $P_x$ coupled to Dzyaloshinskii-Moriya interactions and anisotropic exchange.

Berry-phase calculations yield in-plane polarizations of a few $\mu$C/cm$^2$, which are smaller than those observed in bulk ferroelectrics. However, these are consistent with the polar states induced by stacking in other van der Waals magnets~\cite{ji2023general,chen2025tunable}. Spin-orbit coupling reduces the polarization, reflecting the competition between structural polarity and relativistic interactions that reshape the wavefunction topology and screen the dipole~\cite{feng2016first}. 
Despite this suppression, the predicted polarizations remain within experimental reach by second-harmonic generation~\cite{xin2025nonlinear} and photocurrent spectroscopy~\cite{buscema2015photocurrent}. Thus, $P_x$ provides an experimental fingerprint of stacking registry, linking the nonequivalent iodine environments, enabled by inversion symmetry breaking, to the emergence of anisotropic exchange. The coupling between spin-selective superexchange and inversion breaking offers a microscopic basis for magnetoelectricity in bilayer CrI$_2$, opening pathways for electrical control of antiferromagnetic order and chiral spin textures in reconfigurable 2D spintronic devices~\cite{fert2013skyrmions,gonzalez2025engineering}.

\section{Final Remarks}

This work establishes bilayer CrI$_2$ as a model platform for stacking-controlled magnetism in van der Waals materials. Our systematic first-principles analysis reveals a robust ordering where intralayer exchange dominates the energy scale, stacking symmetry acts as a selector for anisotropic couplings, and weak interlayer exchange provides secondary tunability. The identification of the BA$^\prime$ registry as the ground state, combined with the quantification of exchange parameters across all stable stackings, demonstrates how dimensional engineering strengthens in-plane interactions by $6-10$ \% while preserving the stripe antiferromagnetic ground state.

A central finding is the symmetry-driven activation of anisotropic terms. Dzyaloshinskii-Moriya interactions emerge exclusively in inversion-breaking stackings, establishing a direct structure-property relationship that links registry to chiral exchange. This selectivity, coupled with the registry-dependent in-plane polarization (up to $\sim$10~$\mu$C/cm$^2$), positions bilayer CrI$_2$ as a versatile system where mechanical sliding can simultaneously control magnetic states and electric dipole moments. These magnitudes represent substantial polarization values within the context of stacking-induced ferroelectricity in 2D materials~\cite{ji2023general,chen2025tunable}, indicating robust magnetoelectric coupling optimized for bilayer architectures while remaining modest compared to bulk ferroelectric systems (typically $1-100~\mu$C/cm$^2$). The resulting coupling not only provides a measurable fingerprint of stacking registry but also enables electric-field control of antiferromagnetic order.

From a practical perspective, these results highlight natural pathways for implementation. Defects, folds, or controlled sliding during fabrication can generate indirect stackings, enabling robust noncentrosymmetric phases without external intervention. The moderate energy differences between registries ($25-50$~meV/f.u.) suggest feasible mechanical actuation, while the stability of the exchange hierarchy ensures robustness against thermal fluctuations. This functionality opens pathways for specific device applications: stacking-controlled nonvolatile memory where registry determines both magnetic state and electric polarization, optical detectors sensitive to stacking registry through second-harmonic generation, and mechanically reconfigurable logic gates exploiting the coupled magnetic-electric response. Such capabilities are particularly appealing for mechanically tunable spintronic architectures and moiré superlattices, where stacking can be controlled with nanometer precision.

Our findings contribute to the growing field of stacking engineering in van der Waals magnets~\cite{ji2023general,chen2025tunable,gonzalez2025engineering}, demonstrating how structural control can expand the functional landscape of two-dimensional materials. The comprehensive parameter set established here provides predictive input for finite-temperature magnetism, spin dynamics, and emergent magnetic textures. Altogether, bilayer CrI$_2$ exemplifies the potential of stacking-controlled systems, combining robust exchange interactions, symmetry-selective anisotropy, and registry-dependent magnetoelectricity to create a promising platform for next-generation multifunctional, nonvolatile spintronic and multiferroic devices.

\section*{Acknowledgments}
JWG acknowledges financial support from ANID-FONDECY (Chile) grants N. 1220700 and 1221301.
RAG acknowledges financial support from Basal Program for Centers of Excellence, Grant CIA250002 (CEDENNA).
Powered@NLHPC: This research was partially supported by the supercomputing infrastructure of the NLHPC (CCSS210001).

\section*{Code Availability}
First-principles density functional theory (DFT) calculations were carried out using the open-source \textsc{Quantum ESPRESSO} package, available at \url{https://www.quantum-espresso.org}. Complementary DFT calculations were performed employing the \textsc{OpenMX} code \url{http://www.openmx-square.org/}. For the extraction of pairwise magnetic exchange tensors, we utilized the \textsc{TB2J} postprocessing tool, available at \url{https://github.com/mailhexu/TB2J}.

\bibliography{suya}

%%%%%%%%%%
%%%%%%%%%%
%%%%%%%%%%
\newpage
.
\newpage
\pagebreak
\newpage
\onecolumngrid
\begin{center}

\Large{Supplementary Information}

\large{Stacking-Controlled Magnetic Exchange and Magnetoelectric Coupling in Bilayer CrI$_2$}

\author{B. Valdés-Toro}
\thanks{These authors contributed equally to this work.}
\affiliation{Departamento de Física, Universidad Técnica Federico Santa María, Av. España 1680, Casilla 110V, Valparaíso, Chile}

\author{I. Ferreira-Araya}
\thanks{These authors contributed equally to this work.}
\affiliation{Departamento de Física, Universidad Técnica Federico Santa María, Av. España 1680, Casilla 110V, Valparaíso, Chile}

\author{R. A. Gallardo}
\affiliation{Departamento de Física, Universidad Técnica Federico Santa María, Av. España 1680, Casilla 110V, Valparaíso, Chile}

\author{J. W. Gonz\'alez}
\altaffiliation{Corresponding author:}
\email{jhon.gonzalez@uantof.cl}
\affiliation{Departamento de Física, Universidad de Antofagasta, Av. Angamos 601, Casilla 170, Antofagasta, Chile}

\end{center}

\setcounter{figure}{0} 
\setcounter{section}{0} 
\setcounter{equation}{0}
\setcounter{table}{0}
\setcounter{page}{1}
\renewcommand{\thepage}{S\arabic{page}} 
\renewcommand{\thesection}{S\Roman{section}}   
\renewcommand{\thetable}{S\arabic{table}}  
\renewcommand{\thefigure}{S\arabic{figure}} 
\renewcommand{\theequation}{S\arabic{equation}} 

\section{Antiferromagnetic Orders in Monolayer CrI$_2$}

\begin{figure*}[h!]
\centering
\includegraphics[width=0.8\textwidth]{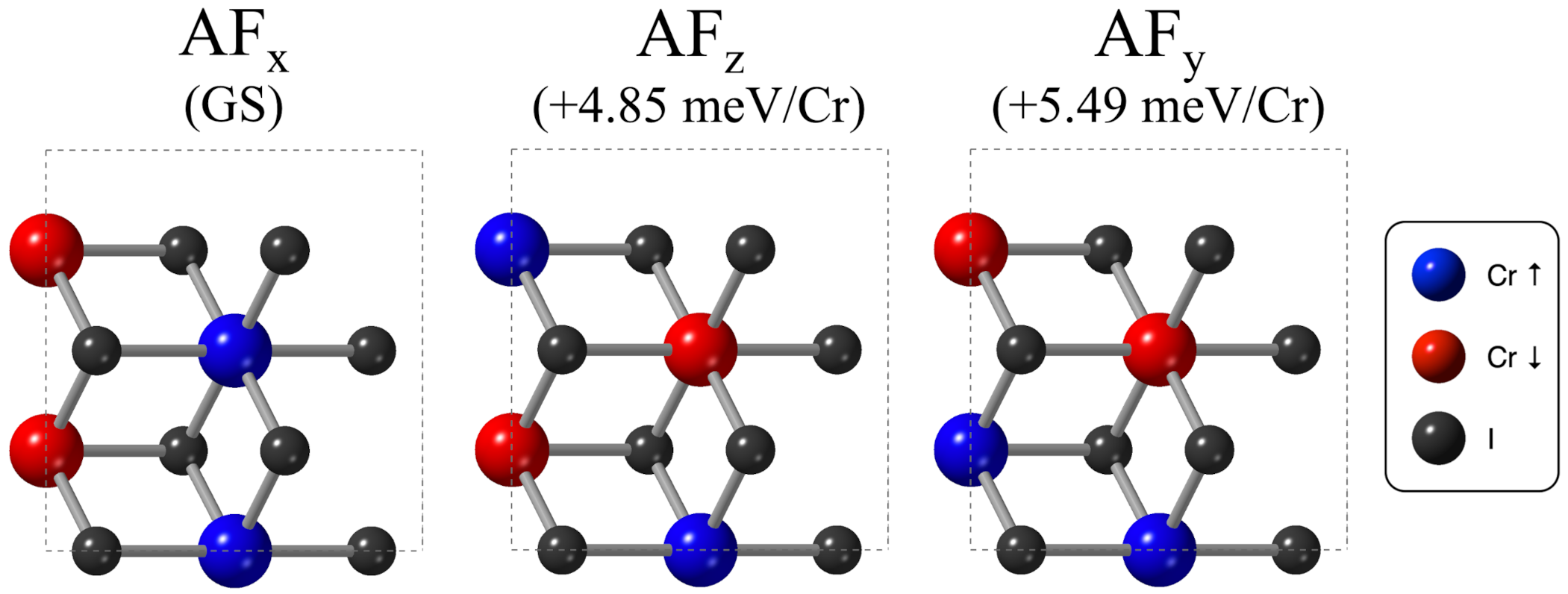}
\caption{
Collinear antiferromagnetic configurations of the CrI$_2$ monolayer in a rectangular  supercell.
The stripe antiferromagnetic ground state AF$_x$ \textbf{(left)}, the zigzag antiferromagnetic configuration AF$_z$ \textbf{(center)}, and the stripe antiferromagnetic configuration AF$_y$ \textbf{(right)}.
Numbers in parentheses indicate the energy difference relative to the AF$_x$ ground state, expressed in meV per Cr atom.
Blue (red) spheres denote Cr atoms with spin up (down), while black spheres represent iodine atoms.
}
\label{fig:s1} 
\end{figure*}

Monolayer CrI$_2$ exhibits three competing collinear antiferromagnetic configurations: the ground state AF$_x$ (striped along $x$), AF$_y$ (striped along $y$), and the zigzag antiferromagnetic configuration AF$_z$, shown in Fig.~\ref{fig:s1}. Magnetic symmetry analysis identifies AF$_x$ as type-I (MSG No.\,56), while both AF$_y$ and AF$_z$ belong to type-IV (MSG No.\,78). DFT calculations reveal a distinct energetic hierarchy, $E_{\mathrm{AF}_x} < E_{\mathrm{AF}_z} < E_{\mathrm{AF}_y}$, with AF$_z$ and AF$_y$ lying $+4.85$\,meV/Cr and $+5.49$\,meV/Cr above the ground state, respectively. 

This stability ordering originates from the interplay between the magnetic topology and the dimerized structure of the Cr sublattice, which features short intra-chain bonds ($d=3.878$\,\AA) and long inter-chain diagonal bonds ($d=4.115$\,\AA). To rationalize the energy differences, we employ a nearest-neighbor Ising Hamiltonian based on the exchange pathways of the AF$_x$ reference state:
\begin{equation}
    \mathcal{H} =
    -J_{s}\!\!\sum_{\langle i,j\rangle_{\rm short}} S_i S_j
    -J_{l}\!\!\sum_{\langle i,j\rangle_{\rm long}} S_i S_j ,
    \label{eq:Ising}
\end{equation}
where $S_i=\pm 1$ represents the spin at site $i$. The exchange parameters extracted from Table~II are $J_s=+2.56$\,meV (ferromagnetic, short bonds) and $J_l=-3.66$\,meV (antiferromagnetic, long bonds).
Within this minimal model, the competition between ferromagnetic short-bond interactions and antiferromagnetic long-bond interactions naturally reproduces the energetic hierarchy $E_{\mathrm{AF}_x} < E_{\mathrm{AF}_z} < E_{\mathrm{AF}_y}$ obtained from the DFT calculations.
This Ising model is intended as an effective description with unit spins; the physical Cr moment ($\sim4\,\mu_B$, corresponding to $S\approx2$) rescales the absolute magnitude of the exchange constants by a factor $1/S^2$, without affecting the relative energy hierarchy discussed here.

\begin{comment}
    
For completeness, we briefly outline the evaluation of the Ising energies for the three collinear AFM configurations. We label the four Cr sites in the primitive cell as $1,2,3,4$, and consider only the nearest-neighbor bonds contained within the cell: two short bonds $(1$--$3)$ and $(2$--$4)$ with coupling $J_s$, and three long bonds $(1$--$2)$, $(2$--$3)$, and $(3$--$4)$ with coupling $J_l$. The corresponding Hamiltonian per unit cell can then be written explicitly as

\begin{equation}
    E = -J_s\left(S_1 S_3 + S_2 S_4\right)
        -J_l\left(S_1 S_2 + S_2 S_3 + S_3 S_4\right),
    \label{eq:Ising-cell}
\end{equation}

where $S_i = \pm 1$ denotes the Ising spin on site $i$.

The three AFM configurations can be represented, up to a global spin flip, as
\begin{align}
    \mathrm{AF}_x &: (S_1,S_2,S_3,S_4) = (+1,-1,+1,-1), \\
    \mathrm{AF}_z &: (S_1,S_2,S_3,S_4) = (+1,-1,-1,+1), \\
    \mathrm{AF}_y &: (S_1,S_2,S_3,S_4) = (+1,+1,-1,-1).
\end{align}

Substituting these spin patterns into Eq.~(\ref{eq:Ising-cell}) yields

\begin{align}
    E_{\mathrm{AF}_x} &= -J_s(1+1) - J_l(-1-1-1) 
    = -2J_s + 3J_l, \\
    E_{\mathrm{AF}_z} &= -J_s(-1-1) - J_l(-1+1-1) 
    = 2J_s + J_l, \\
    E_{\mathrm{AF}_y} &= -J_s(-1-1) - J_l(+1-1+1) 
    = 2J_s - J_l.
\end{align}
%
\end{comment}

\begin{figure*}[!]
\centering
\includegraphics[width=0.4\textwidth]{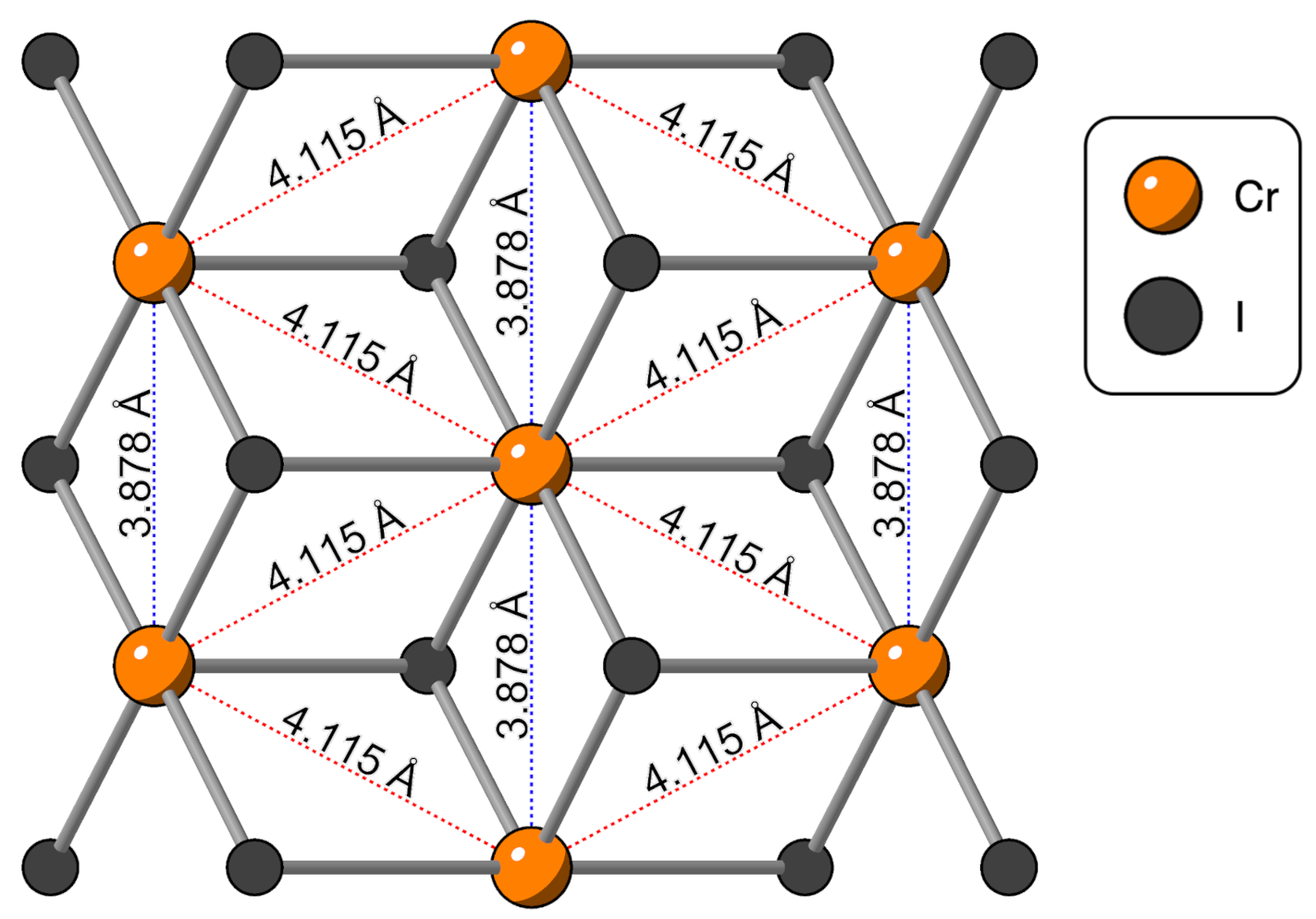}
\caption{
Top view of the CrI$_2$ monolayer showing the inequivalent nearest-neighbor Cr--Cr distances. Blue (red) dotted lines indicate short (long) Cr--Cr bonds of length 3.878~\AA{} (4.115~\AA{}), respectively.
These two distinct bond lengths define the competing ferromagnetic and antiferromagnetic exchange pathways discussed in the main text.
Orange spheres denote Cr atoms and dark gray spheres represent iodine atoms.
}
\label{fig:s2} 
\end{figure*}

\section{Calculation of Magnetic Exchange Parameters using OpenMX and TB2J}

Magnetic exchange interactions are computed using the Green’s-function formalism combined with the magnetic force theorem (MFT), as implemented in the TB2J package~\cite{he2021tb2j}. The magnetic energy is mapped onto a classical Heisenberg Hamiltonian describing bilinear interactions between localized magnetic moments:
\begin{equation}
\label{eq:heisenberg_tb2j}
E = -\sum_{i\neq j}
\left[
J_{ij}^{\text{iso}} \, \vec S_i \cdot \vec S_j
+ \vec S_i \, \mathbf{J}_{ij}^{\text{ani}} \, \vec S_j
+ \vec D_{ij} \cdot (\vec S_i \times \vec S_j)
\right],
\end{equation}
where $\vec S_i$ are normalized spin vectors ($|\vec S_i|=1$), $J_{ij}^{\text{iso}}$ is the isotropic Heisenberg exchange, $\mathbf{J}_{ij}^{\text{ani}}$ is the symmetric anisotropic exchange tensor, and $\vec D_{ij}$ is the Dzyaloshinskii--Moriya interaction (DMI) vector.

This Hamiltonian is equivalent to the tensorial form used internally by TB2J,
\[
E = -\sum_{i\neq j} \vec S_i^{\,T} \mathbf{J}_{ij} \vec S_j ,
\]
with $\mathbf{J}_{ij} = J_{ij}^{\text{iso}}\mathbf{I} + \mathbf{J}_{ij}^{\text{ani}} + \mathbf{A}(\vec D_{ij})$, where $\mathbf{A}(\vec D_{ij})$ is the antisymmetric matrix associated with the DMI.

The calculation proceeds in two steps. First, self-consistent DFT+$U$ calculations are performed using the OpenMX code~\cite{ozaki2003variationally, ohwaki2014method}, which employs a localized pseudo-atomic orbital basis. The Hubbard $U$ correction is applied to the Cr-$3d$ states within the rotationally invariant formalism~\cite{cococcioni2005linear}. Both scalar-relativistic and fully relativistic (including spin-orbit coupling) calculations are performed to generate the Kohn-Sham Hamiltonian and overlap matrices for the reference magnetic configuration.

In the second step, TB2J evaluates the magnetic exchange parameters using the MFT. The method computes intermediate quantities $A_{ij}^{uv}$ ($u,v=x,y,z$), obtained from the Green’s functions of the system, as described in Eqs.~(15)--(17) of Ref.~\cite{he2021tb2j}. These quantities are subsequently combined to yield the isotropic exchange $J_{ij}^{\text{iso}}$, the symmetric anisotropic tensor $\mathbf{J}_{ij}^{\text{ani}}$, and the DMI vector $\vec D_{ij}$.
In particular, the isotropic exchange parameter is given by the Liechtenstein formula~\cite{liechtenstein1987}:
\begin{equation}
\label{eq:liechtenstein}
J_{ij}^{\text{iso}} =
-\frac{1}{4\pi} \, \mathrm{Im}
\int_{-\infty}^{E_F} dE \,
\mathrm{Tr}
\left[
\Delta_i \, \mathbf{G}_{ji}^{\uparrow}(E) \,
\Delta_j \, \mathbf{G}_{ij}^{\downarrow}(E)
\right],
\end{equation}
where $E_F$ is the Fermi energy, $\Delta_i = \mathbf{H}_i^{\downarrow} - \mathbf{H}_i^{\uparrow}$ is the on-site exchange splitting, and $\mathbf{G}_{ij}^{\sigma}(E)$ is the spin-resolved Green’s function between sites $i$ and $j$. The anisotropic exchange and DMI terms are obtained from relativistic generalizations of this expression following the TB2J formalism.

\begin{figure*}[!]
\centering
\includegraphics[width=0.7\textwidth]{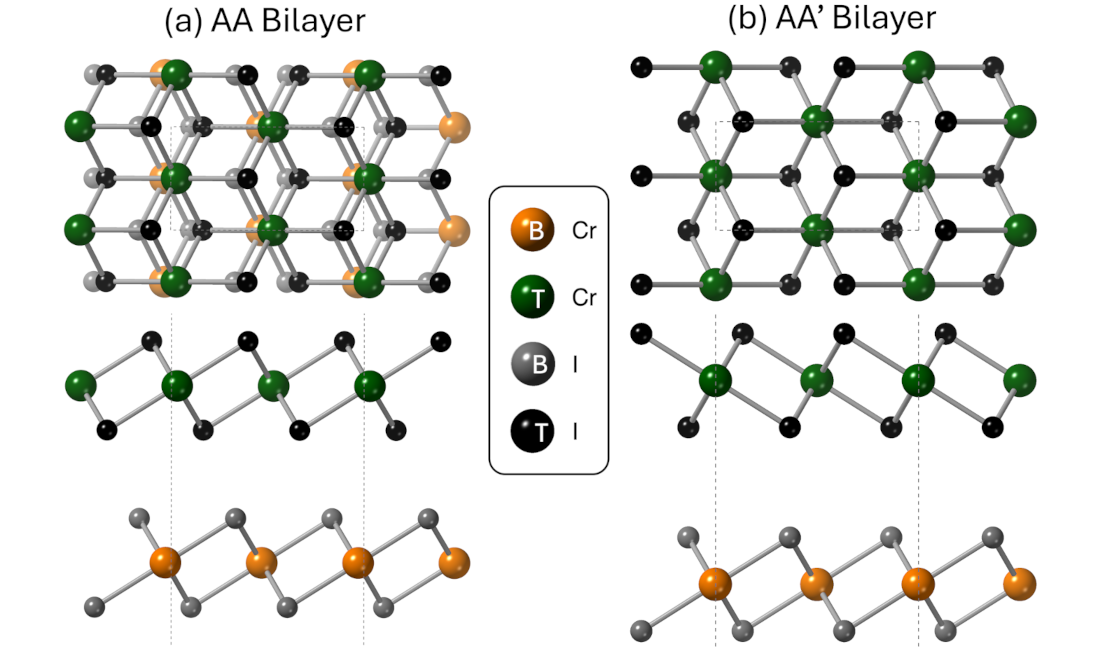}
    \caption{
    Atomic structures of AA-like CrI$_2$ bilayers.
    Top and side views of the \textbf{(a)} direct AA stacking (centrosymmetric) and \textbf{(b)} indirect AA$'$ stacking (non-centrosymmetric), obtained from fully relaxed DFT geometries.
    Chromium atoms in the top (bottom) layer are shown in green (orange); iodine atoms are depicted in dark (light) gray to highlight the distinct vertical alignment of the I sublattices in each configuration.
    }
\label{fig:sa} 
\end{figure*}

\begin{figure*}[!]
\centering
\includegraphics[width=0.7\textwidth]{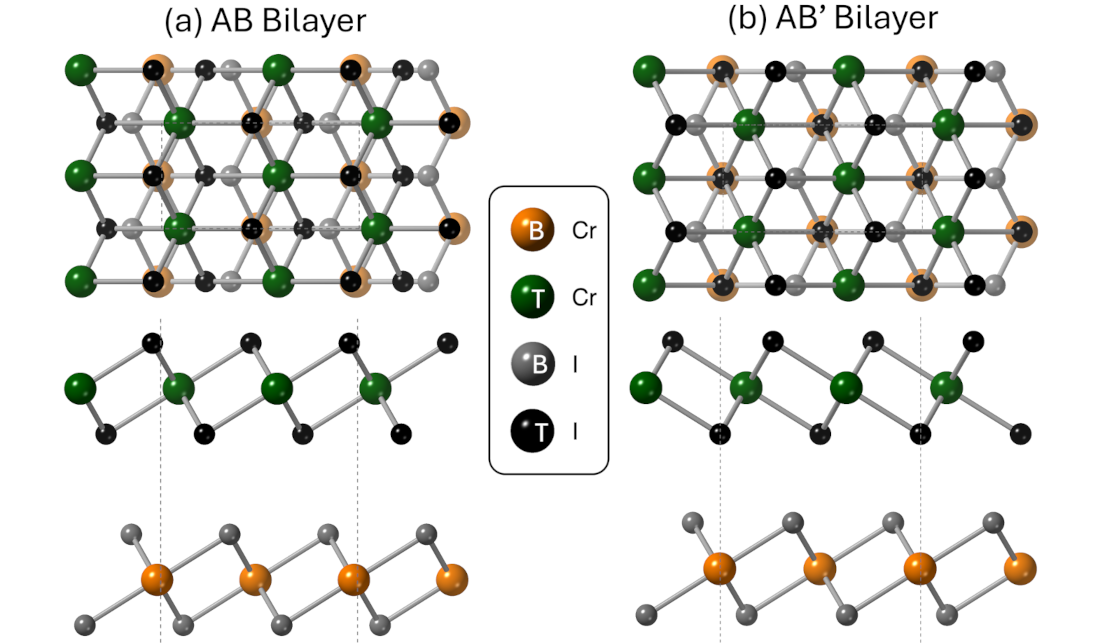}
\caption{
Atomic structures of AB-like CrI$_2$ bilayers.
Top and side views of the \textbf{(a)} direct AB stacking (centrosymmetric) and \textbf{(b)} indirect AB$'$ stacking (non-centrosymmetric), obtained from fully relaxed DFT geometries.
The color scheme follows that of Fig.~\ref{fig:sa}.
}
\label{fig:sb}
\end{figure*}

\begin{figure*}[!]
\centering
\includegraphics[width=0.7\textwidth]{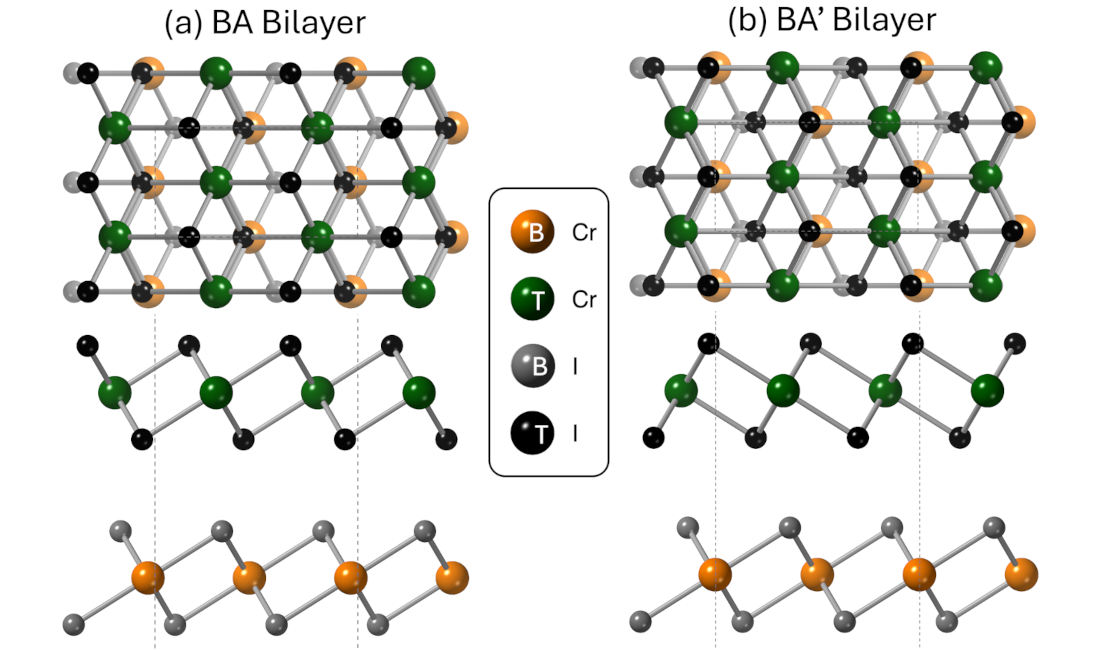}
\caption{
Atomic structures of BA-like CrI$_2$ bilayers.
Top and side views of the \textbf{(a)} direct BA stacking (centrosymmetric) and \textbf{(b)} indirect BA$'$ stacking (non-centrosymmetric), obtained from fully relaxed DFT geometries.
The color scheme follows that of Fig.~\ref{fig:sa}.
}
\label{fig:sc}
\end{figure*}

\section{Symmetry analysis and origin of exchange-induced band splittings in CrI$_2$ bilayers}

The magnetic space groups and crystallographic symmetries of all relaxed CrI$_2$ bilayer configurations were determined using the \texttt{spglib} library (v2.6) with an atomic displacement tolerance of 0.01~\AA. This tolerance reflects a balance between accurately identifying crystallographic symmetries and accommodating minor atomic displacements arising from full structural relaxation. Such a criterion ensures that the assigned symmetries reflect the true physical stacking registry of the relaxed bilayers rather than idealized geometries.

\begin{figure*}[!]
\centering
\includegraphics[width=0.7\textwidth]{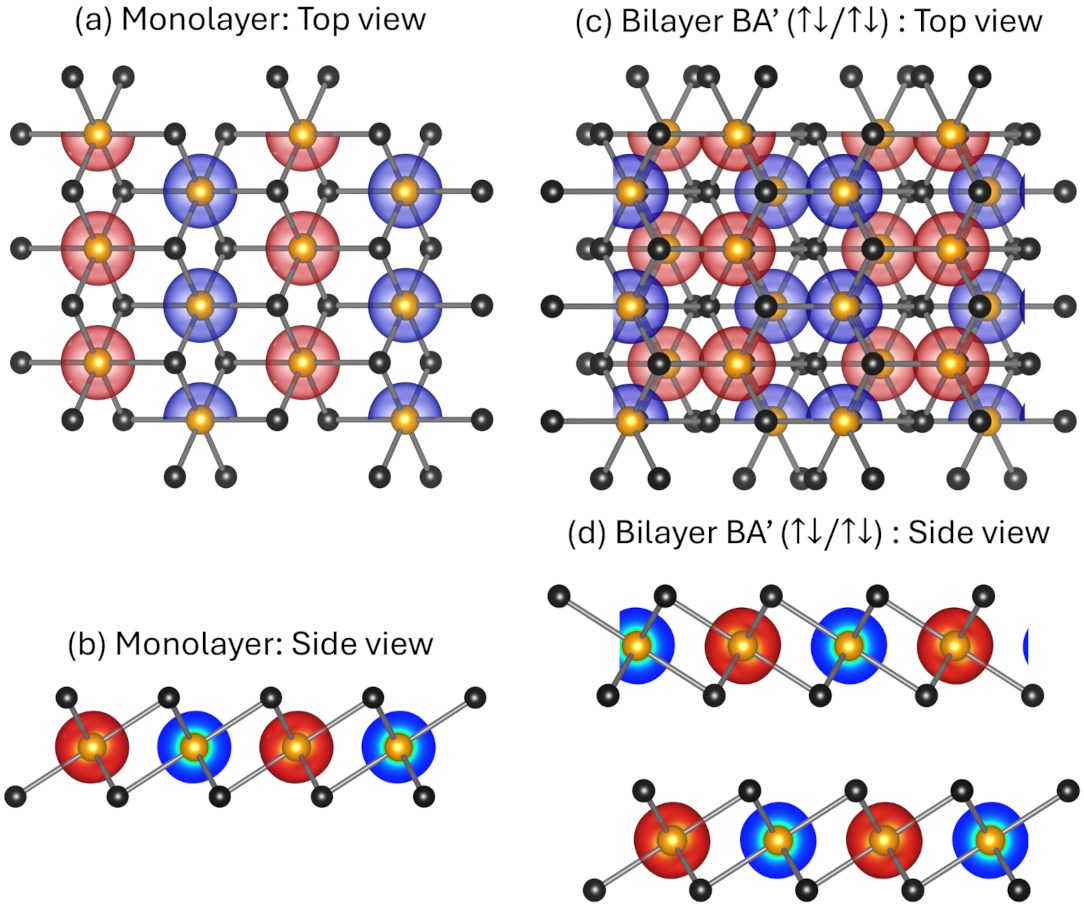}
\caption{
Real-space spin density distribution ($\rho_\uparrow - \rho_\downarrow$) of the CrI$_2$ monolayer in the AF$_x$ configuration and of the BA$^\prime$ bilayer in the antiparallel interlayer configuration ($\uparrow\downarrow/\downarrow\uparrow$).
Panels (a) and (b) show the top and side views of the AF$_x$ monolayer, respectively, while panels (c) and (d) display the corresponding top and side views of the BA$^\prime$ bilayer.
Red and blue isosurfaces correspond to positive and negative spin polarization, respectively, plotted at an isovalue of $0.009$~$e^-$/\AA$^3$. 
}
\label{fig:s3} 
\end{figure*}

The AA-like bilayers shown in Fig.~\ref{fig:sa} are classified as centrosymmetric for the direct AA stacking, belonging to the space group $C2/m$ (No.~12), while the corresponding indirect AA$'$ stacking is non-centrosymmetric and belongs to the polar space group $Amm2$ (No.~38). Similarly, the AB-like bilayers presented in Fig.~\ref{fig:sb} preserve inversion symmetry in the direct AB stacking ($C2/m$, No.~12), whereas the indirect AB$'$ stacking breaks inversion symmetry and is classified within the space group $Cm$ (No.~8). Finally, the BA-like bilayers shown in Fig.~\ref{fig:sc} follow the same symmetry trend: the direct BA stacking remains centrosymmetric and belongs to the space group $C2/m$ (No.~12). In contrast, the indirect BA$'$ stacking lacks inversion symmetry and is assigned to the space group $Pm$ (No.~6).
These symmetry distinctions have direct consequences for the electronic structure of the bilayers. In collinear antiferromagnets, band degeneracy is protected when the system possesses an effective antiunitary symmetry, most commonly the combined $\mathcal{PT}$ operation. More generally, any composite symmetry formed by a spatial operation combined with time reversal can enforce Kramers-like degeneracy throughout the Brillouin zone, even in the absence of spin-orbit coupling.

\begin{figure*}[!]
\centering
\includegraphics[width=0.9\textwidth]{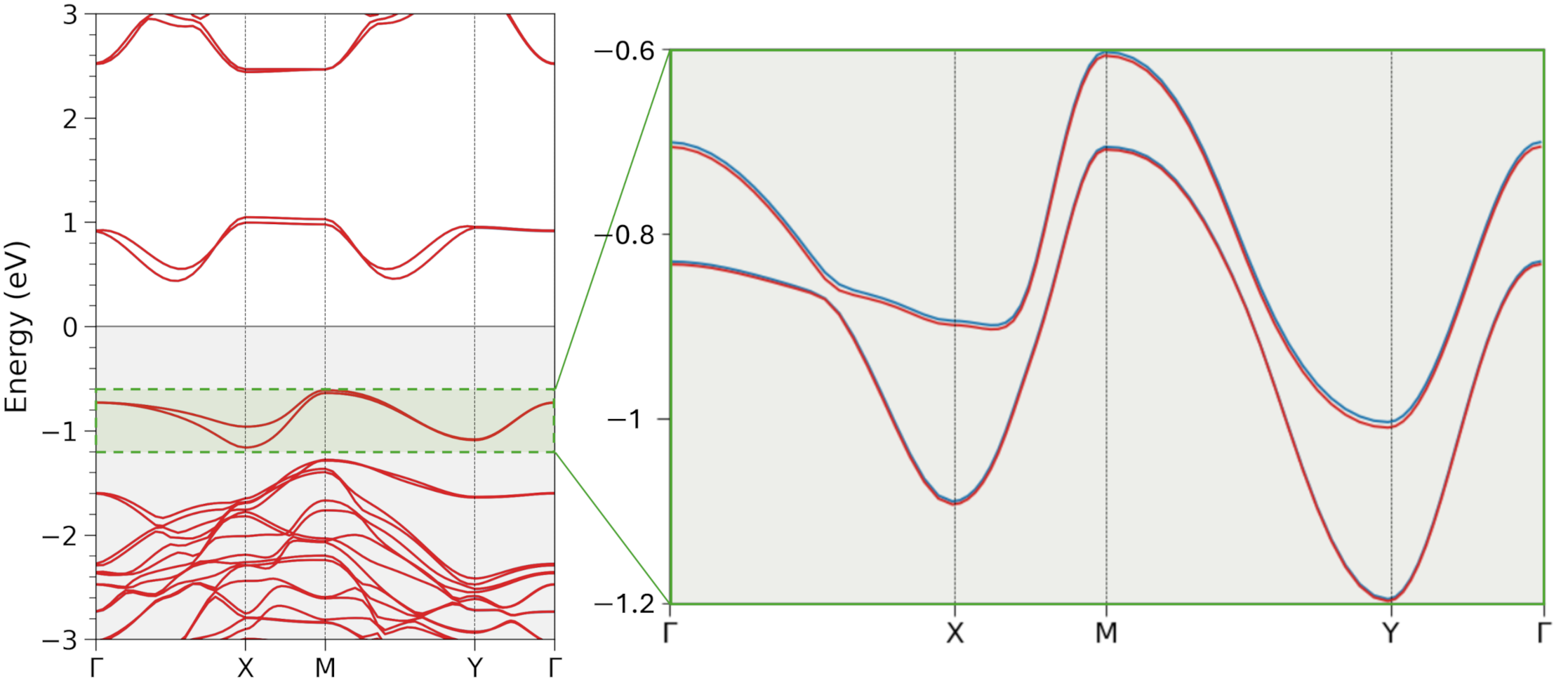}
\caption{
Zoomed-in electronic band structure of the BA$^\prime$ bilayer CrI$_2$ in the antiparallel interlayer configuration ($\uparrow\downarrow/\downarrow\uparrow$).
The band structure shown on the left corresponds to the full dispersion reported in Fig. 3 of main text, while the right panel displays a magnified view of the energy window highlighted by the green shaded region.
Blue and red curves denote bands associated with opposite spin channels, revealing the small exchange-induced splittings near the valence-band maximum. Note that the splittings are on the order of a few meV and are therefore not easily resolved in the full band dispersion. The Fermi level is set to zero energy.
}
\label{fig:s4} 
\end{figure*}

In indirect stackings such as BA$'$, the layer registry breaks global inversion symmetry ($\mathcal{P}$). When this structural asymmetry is combined with a collinear antiferromagnetic order that breaks time-reversal symmetry ($\mathcal{T}$), the protecting $\mathcal{PT}$ symmetry is lost. As a result, the exchange interaction operates within a polar crystal environment, allowing for finite and momentum-dependent spin splittings in the electronic band structure. This mechanism is fully captured at the non-relativistic level and does not rely on spin-orbit coupling, demonstrating that the splittings are purely exchange-driven in origin.
As confirmed by the real-space spin-density distributions shown in Fig.~\ref{fig:s3}, the Cr magnetic moments remain fully compensated and equal in magnitude in both monolayer and bilayer configurations, with only negligible differences at the fourth decimal place. The lifting of spin degeneracy, therefore, does not originate from inequivalent local moments on the two antiferromagnetic sublattices, but from symmetry breaking at the level of the crystal stacking.
The magnitude of these exchange-induced splittings is relatively small (on the order of a few meV), which makes them difficult to resolve in the full band dispersion. To explicitly illustrate this effect, Fig.~\ref{fig:s4} presents a magnified view of the valence-band region for the BA$^\prime$ bilayer (space group $Pm$ (No.~6)) in the antiparallel interlayer configuration ($\uparrow\downarrow/\downarrow\uparrow$). The zoomed-in bands clearly reveal the lifting of spin degeneracy near the valence-band maximum, confirming the presence of exchange-induced splittings.

Taken together, these results demonstrate that the observed band splittings originate from exchange interactions in a non-centrosymmetric antiferromagnetic environment. The symmetry analysis presented above shows that none of the CrI$_2$ bilayer stackings considered in this work realize the non-relativistic spin-group symmetries required for altermagnetism, namely crystallographic operations that map opposite-spin sublattices onto each other while inverting momentum ($\mathbf{k}\rightarrow-\mathbf{k}$). Consequently, the observed splittings arise from exchange physics in a polar antiferromagnet and should not be interpreted as signatures of altermagnetic order.

%%%%

\end{document}